\documentclass[usenatbib]{mn2e}

\usepackage{graphicx}
\usepackage{graphics}
\usepackage{amsmath}
\usepackage{amssymb}
\usepackage{color}
\usepackage{url}
\usepackage{hyperref}
\usepackage{pdflscape}

\def\aap{A\&A}

\def\apj{ApJ}
\def\apjl{ApJL}
\def\apjs{ApJS}

\def\mnras{MNRAS}

\def\araa{ARAA}

\begin{document}

\title[Fast Winds Drive Slow Shells]{Fast Winds Drive Slow Shells: A Model for the Circumgalactic Medium as Galactic Wind-Driven Bubbles}
\author[Lochhaas et al.]{Cassandra Lochhaas$^{1}$, Todd A. Thompson$^{1}$, Eliot Quataert$^{2}$, David H. Weinberg$^{1}$ \\
$^{1}$ Department of Astronomy and the Center for Cosmology and Astroparticle Physics, The Ohio State University, \\
140 West 18th Avenue, Columbus, OH 43210, USA \\
$^{2}$ Astronomy and Physics Departments and Theoretical Astrophysics Center, University of California, Berkeley, \\
Berkeley, CA 94720, USA}

\maketitle

\begin{abstract}
Successful models of the low redshift circumgalactic medium (CGM) must account for (1) a large amount of gas, (2) relatively slow gas velocities, (3) a high degree of metal enrichment, (4) the similar absorption properties around both star-forming and passive galaxies, and (5) the observationally inferred temperature and densities of the CGM gas. We show that galactic wind-driven bubbles can account for these observed properties. We develop a model describing the motion of bubbles driven by a hot, fast galactic wind characteristic of supernova energy injection. The bubble size grows slowly to hundreds of kiloparsecs over $5-10$ Gyr. For high star formation rates or high wind mass loading $\dot M_w/\dot M_\star$, the free-flowing hot wind, the shocked hot wind in the interior of the bubble, and the swept-up halo gas within the bubble shell can all radiatively cool and contribute to low-ionization state metal line absorption. We verify that if the free-flowing wind cools, the shocked wind does as well. We find effective mass loading factors of $(M_w+M_\mathrm{swept})/M_\star\sim5-12$ as the bubbles sweep into the CGM. We predict cool gas masses, velocities, column densities, metal content, and absorption line velocities and linewidths of the bubble for a range of parameter choices. This picture can reproduce many of the COS-Halos and \citet{Keeney2017} observations of low-ionization state metal absorption lines around both star-forming and passive galaxies.
\end{abstract}


%%%%%%%%%%%%%%%%%%%%%%%%%%%%%%%%%%%%%%%%%%%%%%%%%%%%%%%%%%%%%%%%%%%%%%%%%%%%%%%%%%%%%%%%%%%%%%%%%%%
\section{Introduction}
%%%%%%%%%%%%%%%%%%%%%%%%%%%%%%%%%%%%%%%%%%%%%%%%%%%%%%%%%%%%%%%%%%%%%%%%%%%%%%%%%%%%%%%%%%%%%%%%%%%

Observations of galaxy haloes have revealed the ubiquitous presence of multiphase gas. The circumgalactic medium (CGM) extends to hundreds of kiloparsecs and has a range of temperatures $T\sim10^2-10^8$ K and densities $n\sim10^{-6}-1$ cm$^{-3}$ \citep[see references in, e.g.][]{Tumlinson2017}. The hot gas is extended and diffuse and is observed in X-ray emission and absorption lines \citep{Lehnert1999,Strickland2004,Williams2005,Anderson2010,Gupta2012}. The coolest gas in the CGM is dusty and dense and is observed in re-radiated infrared light or molecular line emission \citep{Engelbracht2006,Leroy2015}. The intermediate temperature range of $10^4-10^{5.5}$ K is too cool to emit X-rays and too diffuse to be observed directly in emission lines, so observations are preferentially done in absorption. Such studies probe either the outer CGM through absorption toward a background quasar whose sightline passes through the galaxy's halo \citep{Rudie2012,Tumlinson2013,Bordoloi2014,Keeney2017}, or the inner CGM through ``down the barrel" absorption against the host galaxy's stellar continuum \citep{Heckman2000,Steidel2010,Bordoloi2011,Martin2012,Rubin2014,Chisholm2018}.

Because the CGM is at the interface of accretion inflows that feed galaxies and outflows that are driven by star formation feedback, it is crucial to our understanding of how galaxies and their environments evolve. Galactic winds enrich the CGM and intergalactic medium (IGM) with metals \citep{Aguirre2001,Martin2010,Prochaska2017}, regulate star formation \citep{Murray2005,Ceverino2014,Hopkins2014}, and are necessary to produce observed stellar mass-halo mass relations \citep{Murray2005,Keres2009,Bower2012} and mass-metallicity relations \citep{Tremonti2004,Erb2006,Erb2008,Finlator2008,Peeples2011}. On smaller scales, feedback from star formation in the form of supernovae, massive star winds, and radiation pressure disrupts giant molecular clouds \linebreak \citep{Murray2010,Lopez2011,Murray2011,Raskutti2017}, maintains low star formation efficiencies \citep{Murray2010}, and shapes the interstellar medium (ISM) \citep{McKee1977,Wada2001}. Explaining the diverse CGM observations with galactic winds requires following feedback physics from the size of giant molecular clouds, where winds are launched, all the way to kpc-scale outflows in the CGM, where absorption lines are observed.

The COS-Halos Survey \citep{Werk2012,Tumlinson2013} consists of a sample of 38 quasar sight lines passing within 160 kpc of $z\sim0.2$, $L_\star$ galaxies and focuses on the ionized metal-line absorption present in 33 of the 44 galaxies in the sample. Both low- and high-ionization states are observed, indicating the multi-phase nature of the CGM. The low-ionization metal lines trace cool $\sim10^4-10^5$ K photoionized gas, and the high-ionization metal lines, such as O VI, trace warmer gas at $\sim10^{5.5}-10^6$ K that is likely collisionally ionized \citep[see references in][]{Tumlinson2017}. Large inferred column densities of both metals and neutral hydrogen indicate a massive cool CGM of $\sim3-6\times10^{10}\ M_\odot$ within the virial radius \citep{Peeples2014,Werk2014,Keeney2017}, showing that the CGM is extensive and contains a large fraction of the baryons and metals within the galaxy halo. However, obtaining the physical conditions in the CGM from absorption lines relies on uncertain ionization corrections, photoionizing backgrounds, and temperatures. In addition, most galaxies have only a single quasar sight line passing through their CGM, so trends of density or ionization state with distance from the galaxy can only be determined in a statistical sense from the full sample. Quasar absorption studies suffer from projection effects as well, because absorbing gas may be physically located anywhere along the sight line, not necessarily at a distance equal to the impact parameter. Regardless, physical models must consistently explain the CGM absorption line observations.

There are a number of puzzling features to the COS-Halos Survey observations presented in \citet{Werk2014}. First, the low-ionization state metal absorption arises in cool gas of $T\sim10^4$ K, which cannot be in hydrostatic equilibrium with the hot gaseous halo's thermal gas pressure \citep{Werk2014}. Second, the inferred column densities imply a mass $M\sim10^{10}-10^{11}\ M_\odot$ of cool, metal-enriched gas within the galaxy halo that is comparable to the stellar and interstellar medium mass of the central galaxy \citep{Peeples2014}. Third, the cool gas traced by low-ionization state metals typically has velocity well below the halo escape speed \citep{Werk2014}. Fourth, despite the slow velocities, the cool gas is present hundreds of kiloparsecs from the host galaxy. Fifth, cool gas is common around passive galaxies, indicating that if if originates with star formation, it must be long-lived.

In this paper, we develop a semi-analytic, one-dimensional model for the CGM as a wind-blown bubble driven by feedback from star formation processes. Galactic winds can be driven by supernovae and stellar winds \citep{Creasey2013,Nath2013,vonGlasow2013}, radiation pressure from starlight \citep{Murray2011,Sharma2012,Zhang2012}, cosmic rays \citep{Breitschwerdt1991,Socrates2008,Girichidis2016}, active galactic nuclei (AGN) \citep{Murray2005,Sijacki2007,FaucherGiguere2012,Cicone2014,Harrison2014}, or combinations of any of these \citep{Hopkins2012}. Galactic winds shock and sweep up ambient halo gas as they propagate outward from the host galaxy. Galactic winds and their impact on galaxy evolution have been widely studied in hydrodynamical simulations and semi-analytic models \citep{Samui2008,FaucherGiguere2012,Hopkins2014,Sarkar2015,Fielding2017,Krumholz2017,Nelson2017}. We show that the shocked wind can radiatively cool to $10^4$ K if it is dense, which is the necessary temperature for the low-ionization state metals seen in absorption. Because the galactic wind is driven by feedback from star formation processes, it is enriched in the metals that are byproducts of stellar evolution and can produce large metal column densities. The cooled wind bubble slows to small velocities as it sweeps up CGM gas and slows in the gravitational potential of the galaxy's dark matter halo. However, the wind bubble can still travel to large distances from the host galaxy if the halo gas through which it travels is not too dense, and this process takes a long time. Indeed, as we show in this paper, wind driven bubbles contain $10^{10}M_\odot$ of cool gas, with $v\sim100-300$ km s$^{-1}$, and survive for timescales of $1-10$ Gyr. Under certain conditions and within the context of our simplified models, a galactic wind-blown bubble can qualitatively reproduce all aspects of the COS-Halos observations.

Semi-analytic models are more restricted than hydrodynamic simulations in which processes can be modeled, but can trace the evolution of galactic outflows across a large range of scales and are more easily controlled for exploring the full parameter range of galaxy and wind properties and determining the impact of different physics on the outflow. Previous semi-analytic models of galactic wind bubbles \citep[e.g.,][]{Tegmark1993,Scannapieco2002,Furlanetto2003,Samui2008} follow the analogous stellar wind bubble model of \citet{Castor1975} and \citet{Weaver1977} and focus on predicting the structural properties of the wind bubble and how galaxy and wind parameters affect it. Here, we develop a similar galactic wind bubble model, but develop predictions of observable quantities, like column densities and absorption line properties, and compare the model predictions to recent observations. This model can be used to interpret both observations and results of 3D hydrodynamic simulations.

In \S\ref{sec:model}, we describe the model for the wind-blown bubble, its equations of motion, how we determine if and when the shocked wind can radiatively cool, and the various sets of initial conditions we explore. Section~\ref{sec:shell} presents our results for the physical evolution --- radius, velocity, mass, and thickness --- of the bubble's shell for differing initial conditions, and \S\ref{sec:obs} compares the column densities, temperatures, and absorption lines predicted by our model to the observations and inferred gas properties. Section~\ref{sec:discussion} presents a discussion of our model's results and evaluation of our assumptions. We give a summary in \S\ref{sec:summary}.

%%%%%%%%%%%%%%%%%%%%%%%%%%%%%%%%%%%%%%%%%%%%%%%%%%%%%%%%%%%%%%%%%%%%%%%%%%%%%%%%%%%%%%%%%%%%%%%%%%%
\section{Model}
\label{sec:model}
%%%%%%%%%%%%%%%%%%%%%%%%%%%%%%%%%%%%%%%%%%%%%%%%%%%%%%%%%%%%%%%%%%%%%%%%%%%%%%%%%%%%%%%%%%%%%%%%%%%

\begin{figure}
\includegraphics[width=\linewidth]{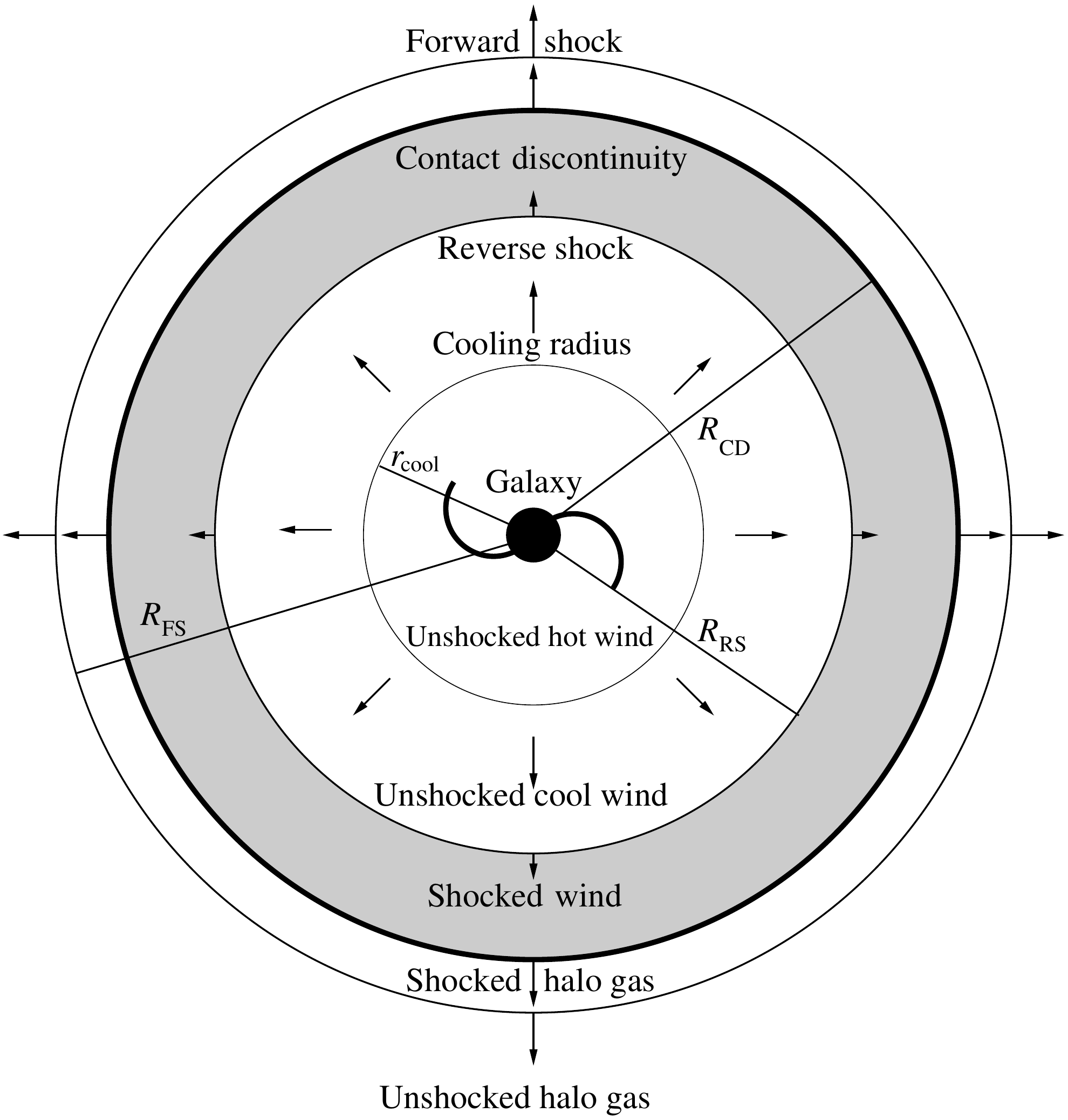}
\caption{Our spherically-symmetric model of the galactic wind-blown bubble. The regions of the bubble structure are unshocked hot wind, unshocked radiatively cooled wind, reverse shock, shocked wind, contact discontinuity, shocked halo gas, and unshocked halo gas, from the centre of the bubble outward. The pressure in the shocked wind region (shaded gray) drives the motion of the bubble.}
\label{fig:cartoon}
\end{figure}

We model a spherically symmetric, hot, and fast wind blowing out of a galaxy with a spherically symmetric gravitational potential provided by a dark matter halo, and model the evolution of the wind as it encounters the surrounding medium numerically. The wind plows into gas in the halo of the galaxy and shocks, creating a forward shock propagating outward. A reverse shock also forms, propagating inward in the frame of the forward shock and shocking the wind. In between the two shocks is a contact discontinuity that separates wind material and swept-up halo gas. The wind bubble thus has the classic structure \citep{Castor1975,Weaver1977,Ostriker1988} with several regions, and we also include cooling of the pre-shock wind beyond its cooling radius \citep{Wang1995,Silich2003,Thompson2016}. The bubble regions are, from inside out: unshocked hot wind, unshocked cool wind, reverse shock, shocked wind, contact discontinuity, shocked halo gas, forward shock, and unshocked halo gas, as shown in Figure~\ref{fig:cartoon}. We do not model the formation of this structure, and instead follow the evolution of the bubble assuming the full structure is already present at the initial bubble radius where we begin our integration. The thermal pressure of the hot, shocked wind between the reverse shock and the contact discontinuity drives the motion of the bubble. We assume no mixing of halo gas and wind material across the contact discontinuity, but we discuss the Rayleigh-Taylor instability in \S\ref{sec:instab}.

%%%%%%%%%%%%%%%%%%%%%%%%%%%%%%%%%%%%%%%%%%%%%%%%%%%%%%%%%%%%%%%%%%%%%%%%%%%%%%%%%%%%%%%%%%%%%%%%%%%
\subsection{Equations of Motion}
\label{sec:eqnsofmotion}

The bubble conserves energy until the shocked wind radiatively cools. The equation of motion is
\begin{equation}
\frac{\mathrm{d}}{\mathrm{d}t}(M_b v_\mathrm{CD}) = 4 \pi R_\mathrm{CD}^2(P_\mathrm{SW}-P_\mathrm{HG})-\frac{GM_bM_\mathrm{enc}(R_\mathrm{CD})}{R_\mathrm{CD}^2} \label{eq:energyconserving}
\end{equation}
where $M_b$ is the mass of the bubble, $v_\mathrm{CD}$ is the velocity of the contact discontinuity, $R_\mathrm{CD}$ is the distance of the contact discontinuity from the galaxy, $P_\mathrm{SW}$ is the thermal pressure of the shocked wind between the reverse shock and the contact discontinuity, $P_\mathrm{HG}$ is the thermal pressure of the (assumed static) halo gas into which the bubble is expanding, and $M_\mathrm{enc}(R_\mathrm{CD})$ is the gravitational mass of the galaxy's dark matter halo enclosed within the radius of the contact discontinuity, $R_\mathrm{CD}$. Note that the shocked wind pressure term is the only positive term on the right hand side of equation~(\ref{eq:energyconserving}) and is therefore responsible for driving the bubble outward.

The mass of the bubble is governed by both the mass deposition of the wind, $\dot M_w$, and the mass swept up by the shell, $\dot M_\mathrm{swept}$:
\begin{align}
\dot M_b&=\dot M_w + \dot M_\mathrm{swept} \\
&= \dot M_w + 4 \pi R_\mathrm{CD}^2 \rho_\mathrm{HG}(R_\mathrm{CD}) v_\mathrm{CD} \label{eq:Mb}
\end{align}
where $\rho_\mathrm{HG}(R_\mathrm{CD})$ is the halo gas density outside the bubble. We explore several $\rho_\mathrm{HG}$ profiles.

The halo gas thermal pressure is given by $P_\mathrm{HG}=n_\mathrm{HG}k_\mathrm{B}T_\mathrm{HG}$, where $n_\mathrm{HG}$ is the number density of the halo gas, $T_\mathrm{HG}$ is the temperature, and $k_\mathrm{B}$ is Boltzmann's constant. We take the halo gas temperature to be $T_\mathrm{HG}\sim10^6$ K, the temperature inferred by observations of X-ray emission and O VII and O VIII absorption in galaxy haloes \citep[see references in][]{Putman2012}. The thermal pressure of the shocked wind is related to the thermal energy of the shocked wind, $E_\mathrm{SW}$, by \citep{FaucherGiguere2012}
\begin{equation}
P_\mathrm{SW}=\frac{2}{3}\frac{E_\mathrm{SW}}{V_\mathrm{SW}}, \label{eq:PSW}
\end{equation}
where $V_\mathrm{SW}$ is the volume of the shocked wind region, $V_\mathrm{SW}=\frac{4\pi}{3}(R_\mathrm{CD}^3-R_\mathrm{RS}^3)$. The differential equation for the thermal energy in the shocked wind is given by
\begin{equation}
\dot E_\mathrm{SW}=L_w - 4\pi R_\mathrm{CD}^2v_\mathrm{CD}\left(P_\mathrm{SW}-P_\mathrm{HG}\right) - L_C \label{eq:dEdt}
\end{equation}
where $L_w=\frac{1}{2}\dot M_w \left[v_w(R_\mathrm{RS})\right]^2$ is the mechanical luminosity of the wind and $v_w(R_\mathrm{RS})$ is the velocity of the wind at the location of the reverse shock, $R_\mathrm{RS}$. The second term on the right hand side in equation~(\ref{eq:dEdt}) describes the energy losses due to work done by the expansion of the bubble against the halo gas, and the third term is energy loss due to radiative cooling:
\begin{equation}
L_C = \frac{4\pi}{3}(R_\mathrm{CD}^3-R_\mathrm{RS}^3)n_\mathrm{SW}^2\Lambda(T_\mathrm{SW}, Z_\mathrm{SW})
\end{equation}
where $n_\mathrm{SW}$ is the number density of the shocked wind, $\Lambda(T_\mathrm{SW}, Z_\mathrm{SW})$ is the volumetric cooling function given by \citet{Wiersma2009} for gas in collisional and photoionization equilibrium at redshift $z=0.2$, and depends on the temperature and metallicity of the shocked wind gas, $T_\mathrm{SW}$ and $Z_\mathrm{SW}$. Note that we do not assume the shocked wind is very metal rich. We assume the wind gas has solar metallicity, $Z_\mathrm{SW}=Z_\odot=0.02$, and the halo gas has $0.1Z_\odot$.

We solve for the location of the reverse shock, $R_\mathrm{RS}$, by assuming the shocked wind is in pressure balance with the unshocked wind's ram pressure. At the reverse shock, the unshocked wind has a ram pressure of
\begin{equation}
P_\mathrm{ram}(R_\mathrm{RS})=\rho_w(R_\mathrm{RS}) v_w(R_\mathrm{RS})^2
\end{equation}
where the wind density is
\begin{equation}
\rho_w(R_\mathrm{RS})=\frac{\dot M_w}{4\pi v_w(R_\mathrm{RS})R_\mathrm{RS}^2} \label{eq:rhowind}
\end{equation}
for a spherical mass-conserving, time-steady outflow. Setting the ram pressure equal to the thermal pressure in the shocked wind region given by equation~(\ref{eq:PSW}) gives
\begin{equation}
\dot M_w v_w(R_\mathrm{RS})(R_\mathrm{CD}^3-R_\mathrm{RS}^3)=2E_\mathrm{SW}R_\mathrm{RS}^2. \label{eq:Rrs}
\end{equation}
Differentiating equation~(\ref{eq:Rrs}) with respect to time gives the equation of motion for $R_\mathrm{RS}$.

The mass deposition rate of the wind is parameterized as
\begin{equation}
\dot M_w=\beta \mathrm{SFR} \label{eq:Mdotw}
\end{equation}
where SFR is the star formation rate and $\beta$ is the dimensionless mass loading factor. The launch velocity of the wind is given by
\begin{equation}
v_l^2=\frac{2\dot E_w}{\dot M_w}
\end{equation}
where $\dot E_w=\alpha \dot E_\mathrm{SN}$ for a supernova-driven wind, $\dot E_\mathrm{SN}$ is the energy output rate of supernovae, and $\alpha$ is a dimensionless parameter that describes the thermalization efficiency of the wind. Assuming $\sim10^{51}$ ergs of energy per supernova explosion and a supernova rate of $\sim(1\ \mathrm{SN\ per\ 100\ years})(\mathrm{SFR}/M_\odot\ \mathrm{yr}^{-1})$, the wind energy injection rate is
\begin{equation}
\dot E_w = 3\times10^{41}\alpha\left(\frac{\mathrm{SFR}}{M_\odot\ \mathrm{yr}^{-1}}\right)\ \mathrm{ergs\ s}^{-1}. \label{eq:Edotw}
\end{equation}
($\dot E_w$ is the energy injection rate of the wind where it is launched, at the galaxy, while $L_w$ (equation~\ref{eq:dEdt}) is the kinetic energy of the free-flowing wind when it impacts the shell at a distance $R_\mathrm{RS}$ from the galaxy.) Note that our model can be applied to any type of wind-driving mechanism, not just supernovae. Other mechanisms will have different forms of $\dot E_w$ than what we give in equation~(\ref{eq:Edotw}), but our overall formalism still holds regardless of the driving process. We can then scale the launch velocity of a supernova-driven wind to
\begin{equation}
v_l=973\left(\frac{\alpha}{\beta}\right)^{1/2}\ \mathrm{km\ s}^{-1}. \label{eq:vl}
\end{equation}
The wind velocity at a distance $r$ from the centre of an NFW halo, assuming the wind reaches its adiabatic terminal velocity at a radius that is small compared to the initial radius of the bubble and then slows down ballistically due to gravity, will be
\begin{equation}
v_w(r)^2=v_l^2 + 8\pi G \rho_\mathrm{0,DM} R_s^3\left[\frac{\log{\left(\frac{r+R_s}{R_s}\right)}}{r} - \frac{\log{\left(\frac{R_l+R_s}{R_s}\right)}}{R_l}\right] \label{eq:vwind}
\end{equation}
where $R_l$ is the launch radius at which $v_w=v_l$, $\rho_\mathrm{0,DM} = M_h/(4\pi R_s^3)\times\left[\log(1+c_v)-c_v/(1+c_v)\right]^{-1}$ is the normalization density of the dark matter halo of mass $M_h$, $R_s=R_\mathrm{vir}/c_v$ is the scale radius and $c_v$ is the concentration of the dark matter halo.

In our fiducial model, the wind blows until $4\times10^9 M_\odot$ of stars are formed, roughly equivalent to the current gas mass reservoir in the Milky Way \citep{Draine2011}. The galaxy thus stops making stars at a time
\begin{equation}
t_\mathrm{SFR}=4\left(\frac{\mathrm{SFR}}{M_\odot\ \mathrm{yr}^{-1}}\right)^{-1}\ \mathrm{Gyr}. \label{eq:tSFR}
\end{equation}
Galaxies with higher SFRs will stop forming stars sooner than galaxies with low SFRs. We explore other star formation histories in \S\ref{sec:SFafter}, but focus on the case where star formation shuts off completely at $t_\mathrm{SFR}$ for the majority of this paper.

Once the star formation shuts off, the wind stops blowing, and the location of the reverse shock can no longer be set by pressure balance between shocked and unshocked wind. Instead, the shocked wind region expands adiabatically at its sound speed, so the reverse shock expands away from the contact discontinuity at the shocked wind sound speed, making its rest-frame radial velocity $v_\mathrm{CD}-c_\mathrm{s,SW}$. If $c_{s,\mathrm{SW}}>v_\mathrm{CD}$, the reverse shock travels backward toward the host galaxy. If $v_\mathrm{CD}$ is sufficiently large, the reverse shock continues traveling outward, away from the host galaxy, even as the shocked wind region expands.

The shocked wind can radiate all its energy if its cooling time is shorter than its advection time. We calculate the cooling time as the total thermal energy in the shocked wind region divided by its cooling rate:
\begin{equation}
t_\mathrm{cool} = \frac{E_\mathrm{SW}}{n_\mathrm{SW}^2\Lambda(T_\mathrm{SW},Z_\mathrm{SW})} = \frac{3}{2}\frac{P_\mathrm{SW}}{n_\mathrm{SW}^2\Lambda(T_\mathrm{SW},Z_\mathrm{SW})}. \label{eq:tcool}
\end{equation}
We assume the density of the shocked wind is constant throughout the shocked wind region:
\begin{equation}
n_\mathrm{SW}=\frac{\dot M_w t}{\frac{4\pi}{3}\mu (R_\mathrm{CD}^3-r_\mathrm{RS}^3)} \label{eq:nSW}
\end{equation}
where $\mu=1.4m_p$ is the mean mass per particle for a typical interstellar medium elemental mixture\footnote{If the gas is fully ionized, $\mu=0.6m_p$ is more appropriate. We take $\mu=1.4m_p$ here to provide a more conservative estimate for the cooling time $t_\mathrm{cool}$.}, $m_p$ is the mass of the proton, and $t$ is the elapsed time the wind has been blowing. The wind shuts off when star formation ends, and the mass in the shocked wind region no longer increases with $t$. Thus, when $t>t_\mathrm{SFR}$, $t\rightarrow t_\mathrm{SFR}$ in equation~(\ref{eq:nSW}). The temperature of the shocked wind, $T_\mathrm{SW}$, is then computed as $T_\mathrm{SW}=P_\mathrm{SW}/(k_\mathrm{B}n_\mathrm{SW})$. This temperature is used in equation~(\ref{eq:tcool}) to compute the cooling time of the shocked wind. The bubble's advection time is
\begin{equation}
t_{\mathrm{adv}}=\frac{R_\mathrm{CD}}{v_\mathrm{CD}}.
\end{equation}
When $t_\mathrm{cool}<t_\mathrm{adv}$, the shocked wind region rapidly cools and we immediately set its temperature to $T_\mathrm{SW}=10^4$ K.

The new equation of motion once the shocked wind cools and becomes momentum-conserving is
\begin{equation}
\frac{\mathrm{d}}{\mathrm{d}t}(M_b v_\mathrm{CD}) = \dot p_w - \frac{GM_bM_\mathrm{enc}(R_\mathrm{CD})}{R_\mathrm{CD}^2} \label{eq:momentumconserving}
\end{equation}
where $\dot p_w = \dot M_w v_w(R_\mathrm{RS})$ is the momentum deposition rate of the wind driving the bubble. The cooling of the shocked wind causes its thermal pressure to drop suddenly, so the location of the reverse shock must adjust to where the ram pressure of the unshocked wind again balances the new, lower thermal pressure of the shocked wind region. We model this adjustment as the unshocked wind ``pushing" the reverse shock at the unshocked wind speed back to a place where pressure balance across the reverse shock is restored, so that $\mathrm{d}R_\mathrm{RS}/\mathrm{d}t=v_w(R_\mathrm{RS})$ until $P_\mathrm{SW}=P_\mathrm{ram}(R_\mathrm{RS})$ again. Because the wind speed is very large, the re-adjustment happens quickly. If the shocked wind has already radiated its thermal energy when the wind turns off at $t=t_\mathrm{SFR}$, the shocked wind's adiabatic expansion back toward the galaxy is slower because its sound speed is smaller.

%%%%%%%%%%%%%%%%%%%%%%%%%%%%%%%%%%%%%%%%%%%%%%%%%%%%%%%%%%%%%%%%%%%%%%%%%%%%%%%%%%%%%%%%%%%%%%%%%%%
\subsection{Fiducial Initial Conditions}
\label{sec:init}

Our fiducial galaxy has an NFW \citep{Navarro1996} dark matter halo density profile with a halo mass $M_h=10^{12}\ M_\odot$, concentration $c_v=13$ \citep{Maller2004}, and virial radius $R_\mathrm{vir}=320$ kpc, equal to the median virial radius of galaxies in the COS-Halos sample estimated by \citet{Tumlinson2013}. We launch the bubble from a radius $R_l=1$ kpc at $t=0$ and follow its evolution for 100 Gyr, or until it falls back to the host galaxy. The initial radius of the reverse shock is set to 90\% of the initial radius of the contact discontinuity, $R_l$. The bubble is launched at a low velocity relative to the wind, $100$ km s$^{-1}$, but approaches the initial wind speed, $v_l$, within the first few time steps of integration. The initial mass and thermal energy of the shell are set to zero.

We explore three SFRs, four halo gas density profiles, and three values of the mass loading factor of the wind with respect to the star formation rate, $\beta$. Table~\ref{tab:models} lists the combinations of $\beta$ and SFR we use in our fiducial models, and the effect of these parameters on the wind mass outflow rate, launch velocity of the wind, and star formation shut off time (equation~\ref{eq:tSFR}). The galaxy forms $4\times10^9\ M_\odot$ of stars in every model, but models with varying $\beta$ have different mass outflow rates (see the third column in Table~\ref{tab:models}), and hence varying reservoirs of gas. The three SFRs are $1,\ 10,$ and $100\ M_\odot$ yr$^{-1}$, which place the star formation cutoff time at $t_\mathrm{SFR}=4, 0.4,$ and $0.04$ Gyr, respectively. The three values of $\beta$ we consider are $\beta=1,\ 2,$ and 4, which place the wind launch velocity at $v_l=973,\ 688,$ and $486$ km s$^{-1}$, respectively (equation~\ref{eq:vl}). Two of the halo gas density profiles we examine are power law profiles,
\begin{equation}
\rho_\mathrm{HG}=\rho_0\left(\frac{r}{R_0}\right)^{-x},
\end{equation}
which are all normalized to a density $\rho_0=10^{-3}\mu$ cm$^{-3}$ ($\mu=1.4m_p$) at the normalization radius $R_0=10$ kpc, roughly in line with the $n_\mathrm{H}(R)$ profile found by \citet{Werk2014}. We use powers of $x= 1$ or $2$. We also examine an NFW halo gas profile with the same normalization density and scale radius $R_0$, and a Maller \& Bullock gas profile \citep[][hereafter referred to as ``MB"]{Maller2004}. All density profiles have the same ``core" such that $\rho_\mathrm{HG}(r<R_0)=\rho_0$. Denser halo gas profiles will slow down the evolution of the bubble as more mass piles onto the contact discontinuity, while less dense halo gas profiles may allow the expanding wind shell to escape to large radius.

The first two columns in Table~\ref{tab:den} list the halo gas density profiles and their corresponding integrated gas mass within $R_\mathrm{vir}$. The halo gas density profiles and masses listed are the initial conditions, before feedback launches winds into the halo. Feedback supplies an additional $4\times10^9\ M_\odot$, $8\times10^9\ M_\odot$, or $1.6\times10^{10}\ M_\odot$ of gas to the CGM by the time star formation ends at $t_\mathrm{SFR}$, for $\beta$ values of 1, 2, or 4, respectively. The last two columns in Table~\ref{tab:den} list the range of ambient + wind gas masses in the halo after $t_\mathrm{SFR}$, for $\beta=1$ through 4, and the corresponding range of gas baryon fractions. There is a large range of halo gas masses across the fiducial models, and we caution that because some of the $\rho_\mathrm{HG}$ functions produce far too much or too little halo gas, it is unlikely that any one of the density profiles describes the halo gas pre-feedback completely. The halo gas density profiles that provide gas masses closest to gas fractions from cosmological simulations \citep[e.g.,][]{vandeVoort2016} when combined with galactic wind mass are the power-law profile with $x=2$ and the MB profile, but our models of these profiles have gas baryon fractions roughly four times smaller than expected for a $10^{12}M_\odot$ dark matter halo from simulations. We model a single burst of star formation, and do not account for cosmological filamentary feeding of gas onto the galaxy; such processes could provide additional gas to the halo to increase the gas baryon fraction. Therefore, our results spanning the varied models should be taken as a range of possible wind bubble outcomes, from one extreme to another, from a single star formation epoch.

\begin{table*}
\begin{minipage}{175mm}
\centering
\begin{tabular}{c r r r r}
$\beta$ & \multicolumn{1}{c}{SFR ($M_\odot$ yr$^{-1}$)} & \multicolumn{1}{c}{$\dot M_w$ ($M_\odot$ yr$^{-1}$)} & \multicolumn{1}{c}{$v_l$ (km s$^{-1}$)} & \multicolumn{1}{c}{$t_\mathrm{SFR}$ (Gyr)} \\
\hline
1 & 1 & 1 & 973 & 4 \\
1 & 10 & 10 & 973 & 0.4 \\
1 & 100 & 100 & 973 & 0.04 \\
2 & 1 & 2 & 688 & 4 \\
2 & 10 & 20 & 688 & 0.4 \\
2 & 100 & 200 & 688 & 0.04 \\
4 & 1 & 4 & 486 & 4 \\
4 & 10 & 40 & 486 & 0.4 \\
4 & 100 & 400 & 486 & 0.04 \\
\end{tabular}
\caption{The values of $\beta$ and SFR we vary in our fiducial models, and the corresponding wind outflow rate $\dot M_w$ (equation~\ref{eq:Mdotw}), wind launch velocity $v_l$ (equation~\ref{eq:vl}), and star formation shut off time $t_\mathrm{SFR}$ (equation~\ref{eq:tSFR}). For each combination of $\beta$ and SFR listed here, we explore four halo gas density profiles (see Table~\ref{tab:den}).}
\label{tab:models}
\end{minipage}
\end{table*}

\begin{table*}
\begin{minipage}{175mm}
\centering
\begin{tabular}{c c c c}
Halo gas density profile & $M_\mathrm{HG}$ ($M_\odot$) & Ambient + wind gas mass after $t_\mathrm{SFR}$ ($M_\odot$) & Halo gas baryon fraction after $t_\mathrm{SFR}$ \\
\hline
$\propto r^{-1}$ & $2.2\times10^{11}$ & $2.3\times10^{11}-2.4\times10^{11}$ & $0.226-0.238$ \\
$\propto r^{-2}$ & $1.4\times10^{10}$ & $1.8\times10^{10}-3.0\times10^{10}$ & $0.018-0.030$ \\
NFW & $4.2\times10^9$ & $8.2\times10^{9}-2.0\times10^{10}$ & $0.008-0.020$ \\
MB & $2.0\times10^{10}$ & $2.4\times10^{10}-3.6\times10^{10}$ & $0.024-0.036$ \\
\end{tabular}
\caption{The halo gas density profiles $\rho_\mathrm{HG}$ we use in our fiducial models, the corresponding integrated halo gas mass within $R_\mathrm{vir}$ before feedback turns on, $M_\mathrm{HG}$, the range of ambient + wind gas masses within the halo after star formation has finished for the range of $\beta$ values, and the corresponding baryon fractions of the halo gas after star formation has finished.}
\label{tab:den}
\end{minipage}
\end{table*}

%%%%%%%%%%%%%%%%%%%%%%%%%%%%%%%%%%%%%%%%%%%%%%%%%%%%%%%%%%%%%%%%%%%%%%%%%%%%%%%%%%%%%%%%%%%%%%%%%%%
\subsection{Caveats to the Model}
%%%%%%%%%%%%%%%%%%%%%%%%%%%%%%%%%%%%%%%%%%%%%%%%%%%%%%%%%%%%%%%%%%%%%%%%%%%%%%%%%%%%%%%%%%%%%%%%%%%

While simple, computationally inexpensive models are useful for physical understanding and parameter space exploration, there are several drawbacks to the model presented in this paper. The gravitational potential and launch structure of the wind are both spherically symmetric because we model in one dimension, so we do not include the potential of the galactic disk or the biconical outflow shape often observed and modeled for galactic winds. However, because we model several halo gas density profiles, a combination of several bubbles expanding into different density profiles could represent wind expanding in different directions in a non-spherically-symmetric halo gas density profile. In that case, one density profile could represent a wind bubble expanding along the minor axis out of a galactic disk, while a separate density profile would represent the bubble expanding at an angle to the minor axis, for example.

We also use only smoothly monotonic density profiles for the halo gas, so we do not allow for clumps or clouds in the unshocked ambient halo gas. If the shell impacts clumps, it would slow down, while parts of the shell that do not impact clouds would continue, possibly fragmenting the shell. If the shocked wind bounded by the contact discontinuity and reverse shock had not yet cooled, it could escape through holes in the shell and the dynamics of the bubble would change as the wind energy leaks out, just as in the case of shell fragmentation due to instabilities (\S\ref{sec:instab}) \citep{HarperClark2009,Rogers2013,Zubovas2014}.

The halo gas density profiles we use also do not allow for large-scale variations over the full sphere surrounding the galaxy, such as dense filaments of cold gas that are expected to provide gas to star-forming galaxies \citep[e.g.,][]{Keres2005}. By exploring a range of halo gas density profiles, we hope to capture the effect of wind bubbles traversing a range of gas structures in the halo. \citet{ElBadry2018} found a spherically-averaged cumulative mass distribution that is roughly proportional to radius, outside of $0.1R_\mathrm{vir}$, surrounding galaxies in the FIRE simulations with $10^{12}M_\odot$ haloes. This implies a halo gas density, on average, that is roughly proportional to $r^{-2}$.

Other than the ambient halo gas density profile, the star formation history of the galaxy is the most important factor in determining the motion of the bubble. We assume simple star formation histories of constant SFR that is instantaneously shut off at $t_\mathrm{SFR}$. Galaxy simulations show that realistic star formation histories may be more complex than we assume, with periods of bursts and quiescence \citep{Sparre2017,FaucherGiguere2018}, and inflowing gas replenishing the star formation reservoir. We also assume the mass of the galaxy remains constant over the full time we track the bubble expansion, which is not true for real galaxies. The mass of the galaxy affects the wind bubble only through its gravity, which, along with the halo gas density profile, affects the slow-down of the bubble and if and when it eventually falls back to the galaxy. The star formation history has a stronger effect on the bubble, as the energy and momentum input by star formation feedback in the form of supernovae are the only driving force of the wind bubble. As evidenced by our differing results with differing SFRs, a single bubble that experiences many star formation episodes driving it at different times may have different and more complicated dynamics than we model here. However, the case of large SFR over a short period of time, like our models with SFR $=100\ M_\odot$ yr$^{-1}$ that ends at $t_\mathrm{SFR}=0.04$ Gyr, could represent a short-lived but intense starburst.

%%%%%%%%%%%%%%%%%%%%%%%%%%%%%%%%%%%%%%%%%%%%%%%%%%%%%%%%%%%%%%%%%%%%%%%%%%%%%%%%%%%%%%%%%%%%%%%%%%%
\section{Evolution of the Shell}
\label{sec:shell}
%%%%%%%%%%%%%%%%%%%%%%%%%%%%%%%%%%%%%%%%%%%%%%%%%%%%%%%%%%%%%%%%%%%%%%%%%%%%%%%%%%%%%%%%%%%%%%%%%%%

\begin{landscape}
\begin{figure}
\centering
\includegraphics[width=\linewidth]{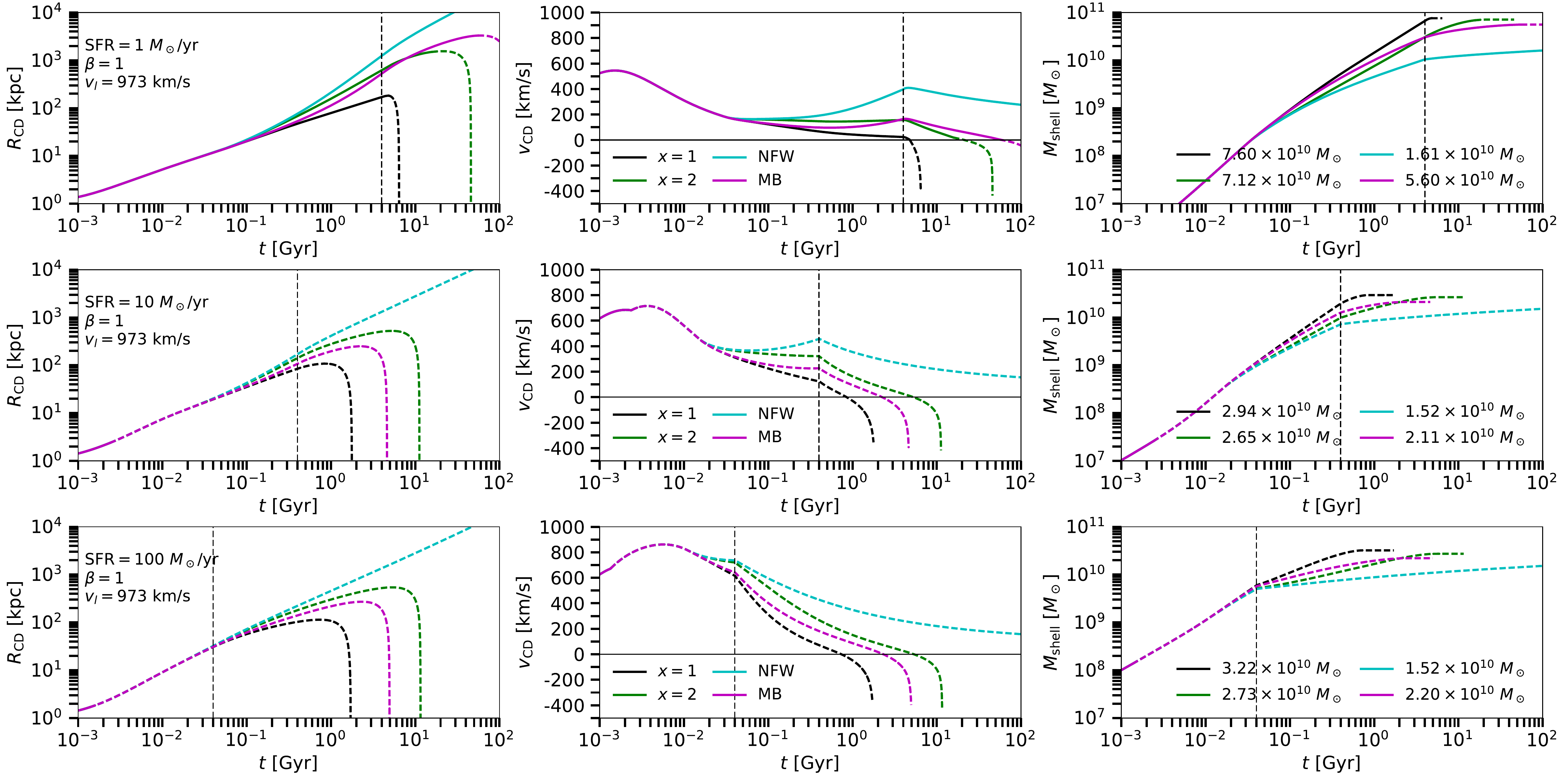}
\caption{The radius (left panels) and velocity (middle panels) of the contact discontinuity and mass of the bubble (right panels) as functions of time, all for wind bubbles run with $\beta=1$. Curve color indicates runs with different halo gas density profiles; black and green curves for power-law halo gas density profiles with powers $x=1$ and 2, cyan and magenta curves for NFW and MB halo gas density profiles. From top to bottom in left and centre panels, the profiles are NFW, power-law with $x=2$, MB, then power-law with $x=1$. This order is flipped in right panels. Top row shows bubbles driven by a SFR of $1\ M_\odot$ yr$^{-1}$, middle row shows those driven by a SFR of $10\ M_\odot$ yr$^{-1}$, and bottom row shows those driven by $100\ M_\odot$ yr$^{-1}$. The curves are solid when the bubble is energy-conserving and switch to dashed when the shocked wind radiatively cools and the bubble's evolution switches to momentum-conserving. Vertical dashed lines in each panel show the time when star formation turns off, at $4,\ 0.4,$ or $0.04$ Gyr, respectively, from top to bottom. Legends in right panels show the maximum mass of the shell, which occurs either when it falls back to the galaxy or at the end of integration. All bubbles shown here were launched from a radius of $R_l=1$ kpc with a wind launch speed of $v_l=973$ km s$^{-1}$. See online journal for a color version of this figure.}
\label{fig:Rv}
\end{figure}
\end{landscape}

\begin{landscape}
\begin{figure}
\centering
\includegraphics[width=\linewidth]{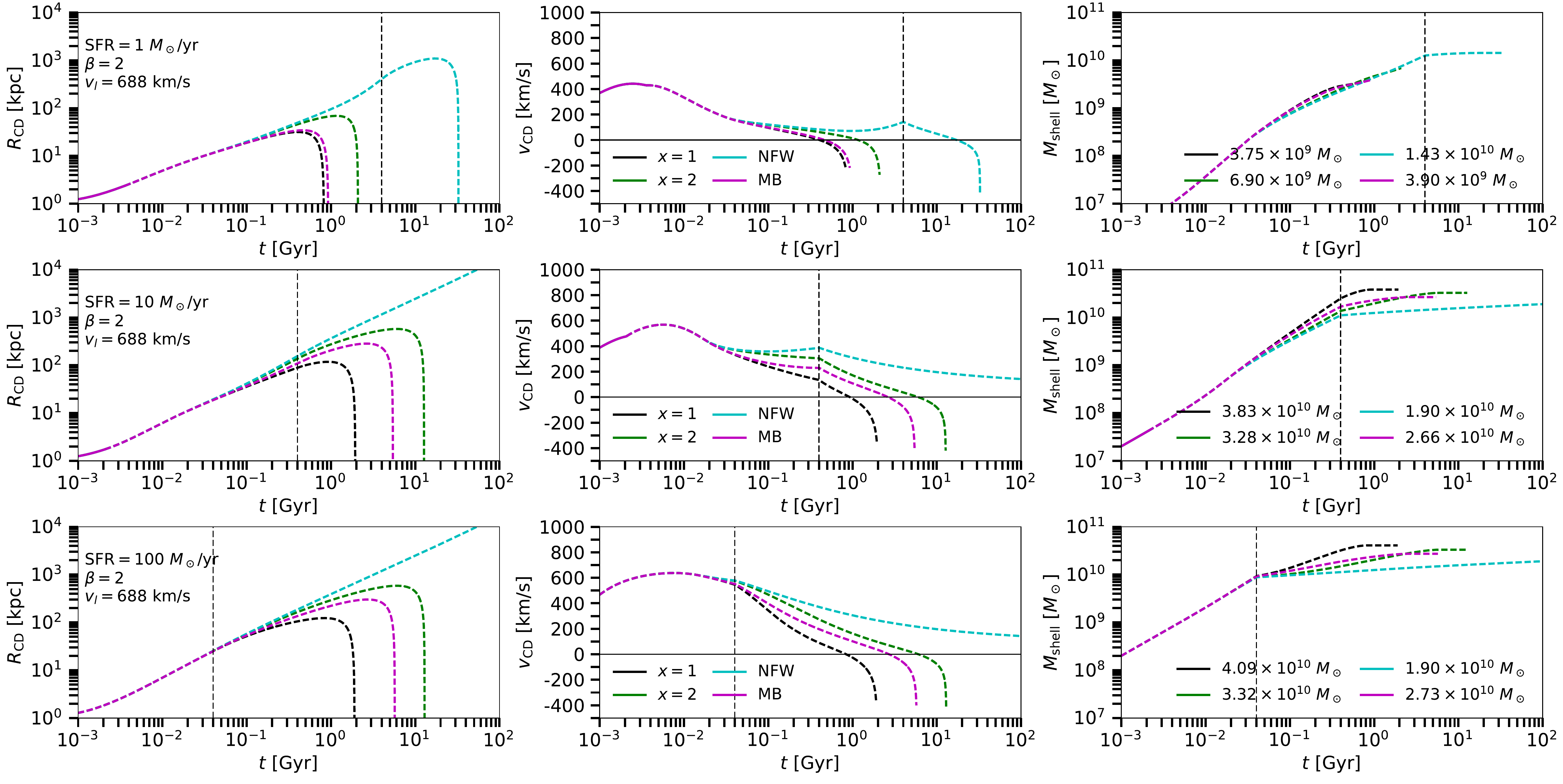}
\caption{Same as Figure~\ref{fig:Rv}, but for $\beta=2$. In these cases, the wind launch velocity is $v_l=688$ km s$^{-1}$ (equation~\ref{eq:vl}).}
\label{fig:Rv_beta2}
\end{figure}
\end{landscape}

\begin{landscape}
\begin{figure}
\centering
\includegraphics[width=\linewidth]{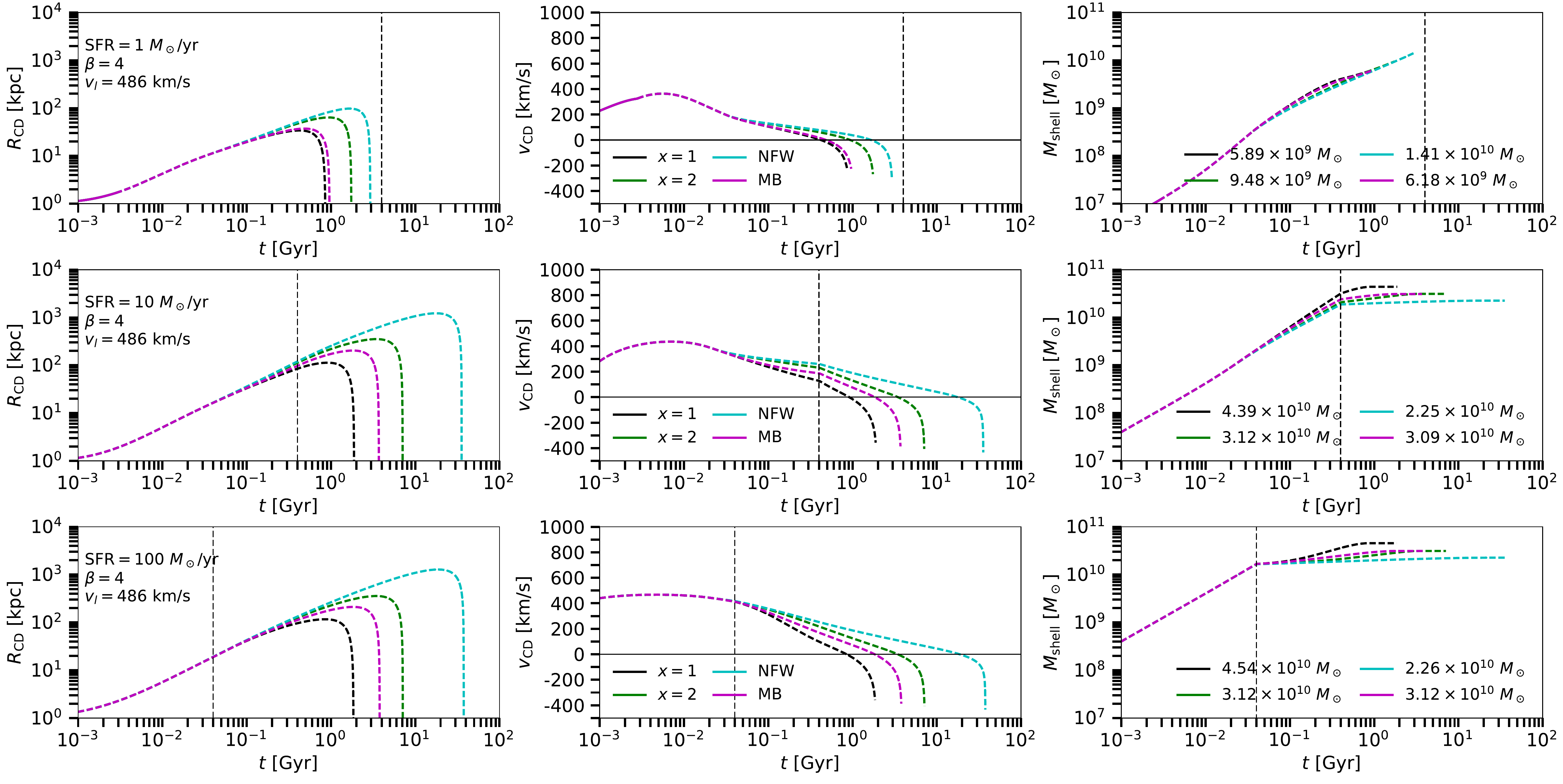}
\caption{Same as Figure~\ref{fig:Rv}, but for $\beta=4$. In these cases, the wind launch velocity is $v_l=486$ km s$^{-1}$ (equation~\ref{eq:vl}).}
\label{fig:Rv_beta4}
\end{figure}
\end{landscape}

We have defined a system of four coupled differential equations given by equation~(\ref{eq:energyconserving}) (or equation~\ref{eq:momentumconserving} after the shocked wind has cooled), equation~(\ref{eq:Mb}), equation~(\ref{eq:dEdt}), and the time derivative of equation~(\ref{eq:Rrs}). We use a standard Runge-Kutta fourth-order numerical integrator with adaptive time steps to compute the contact discontinuity radius, velocity, total shell mass, thermal energy of the bubble, and the reverse shock radius. All runs start out with the energy-conserving equation of motion (equation~\ref{eq:energyconserving}) until the shocked wind radiatively cools, then the calculation switches to the momentum-conserving equation of motion (equation~\ref{eq:momentumconserving}). Figures~\ref{fig:Rv},~\ref{fig:Rv_beta2}, and~\ref{fig:Rv_beta4} show the radius (left panels), velocity (middle panels), and mass (right panels) of the shell as functions of time for all runs with $\beta=1,\ 2,$ or 4, respectively; each row shows a different SFR, and each differently colored curve in each panel shows a different halo gas density profile. Solid curves indicate where the bubble is still in the energy-conserving phase, before the shocked wind cools. Where the curves switch to dashed lines indicates that the shocked wind has radiatively cooled and the bubble is now momentum-conserving. Vertical black dashed lines show the time at which star formation shuts off, $t_\mathrm{SFR}$. The legends in the right panels indicate the mass of the shell at either the time when the shell falls back to the galaxy, if it does, or at the end of integration, if it does not. We integrate all bubble evolution runs until 100 Gyr to capture late-time fall-back, but focus on the evolution of the bubble only within the galaxy halo because we do not model the transition from halo to intergalactic medium.

The majority of the bubbles reach a mass of both shocked wind material and swept-up halo gas of $\sim2-5\times10^{10}\ M_\odot$ by either the time they fall back to the galaxy, or the integration ends, as shown by the legends in the right panels. The most massive bubbles are those with $\beta=1$, SFR $=1\ M_\odot$ yr$^{-1}$, and expanding through power-law halo gas density profiles, but these bubbles do not have radiatively cooling shocked wind until after they begin to fall back toward the galaxy. The total mass of the bubble depends on both the mass of wind material passing through the reverse shock and the integrated halo gas density profile out to the

\noindent maximum radius the bubble reaches during the integration time. Because there is a non-monotonic relation between bubbles with higher wind mass (larger $\beta$ or larger SFR) and how far they travel, there is also a non-monotonic relation between $\beta$ or SFR and the total bubble mass at the end of integration. For example, the bubble with SFR $=10\ M_\odot$ yr$^{-1}$ expanding through a power-law halo gas density profile with power $x=2$ (red curves in centre panels in Figures~\ref{fig:Rv},~\ref{fig:Rv_beta2}, and~\ref{fig:Rv_beta4}) reaches $\sim500$ kpc when $\beta=1$, $\sim600$ kpc when $\beta=2$, but $\sim300$ kpc when $\beta=4$. Accordingly, the total masses of these bubbles when they fall back to the galaxy are $2.65\times10^{10}\ M_\odot$ when $\beta=1$, $3.28\times10^{10}\ M_\odot$ when $\beta=2$, and $3.12\times10^{10}\ M_\odot$ when $\beta=4$. The change in the bubble mass is not proportional to the change in $\beta$.

For all SFRs and values of $\beta$, the NFW (cyan) halo gas density profile allows bubbles to travel the furthest during the integration time and also tends to produce the least massive bubbles. For most values of $\beta<4$, these bubbles escape the host galaxy, driving a wind that will never fall back. The exception is the case with $\beta=2$ and SFR $=1\ M_\odot$ yr$^{-1}$, which has a bubble mass at the time of fall-back of $\sim1.4\times10^{10}\ M_\odot$, larger than bubbles run with the other halo gas density profiles for this combination of $\beta$ and SFR. This is because this bubble travels significantly further than those traveling through other halo gas density profiles, so despite the steeply-declining NFW density profile, it sweeps up more mass overall. As the middle panels show, the velocities of the shells moving through the NFW halo gas density profile are much higher than the other density profiles.

For those bubbles that collapse after $t_\mathrm{SFR}$ when the wind is no longer blowing, the collapse is free-fall. If fall back occurs before $t_\mathrm{SFR}$, momentum input from the still-blowing wind can slow the fall somewhat, but the wind momentum is not large enough to overcome the gravitational force. All the swept-up mass and wind mass returns to the galaxy, and the time from shell turn-around to galaxy return can be between $\sim300$ Myr and $\sim10$ Gyr, where shells that have traveled further before falling take longer to return to the galaxy. Most bubbles that fall have a total (swept + wind) mass of $M_b\sim$few $\times 10^{10}\ M_\odot$, which could trigger another episode of star formation. However, our assumption of simple free-fall is likely inaccurate, as any fragmentation of the shell as it travels through the halo could populate the interior volume of the bubble with clouds that alter the motion of the shell as it falls. In addition, radiation pressure from the galaxy's light \citep{Murray2005,Murray2011}, ongoing stellar winds, or other processes we do not model, could change the properties of shell as it falls. We focus here on the properties of the shell while it is in the CGM and not the effects of its return to the galaxy.

\begin{figure}
\centering
\includegraphics[width=\linewidth]{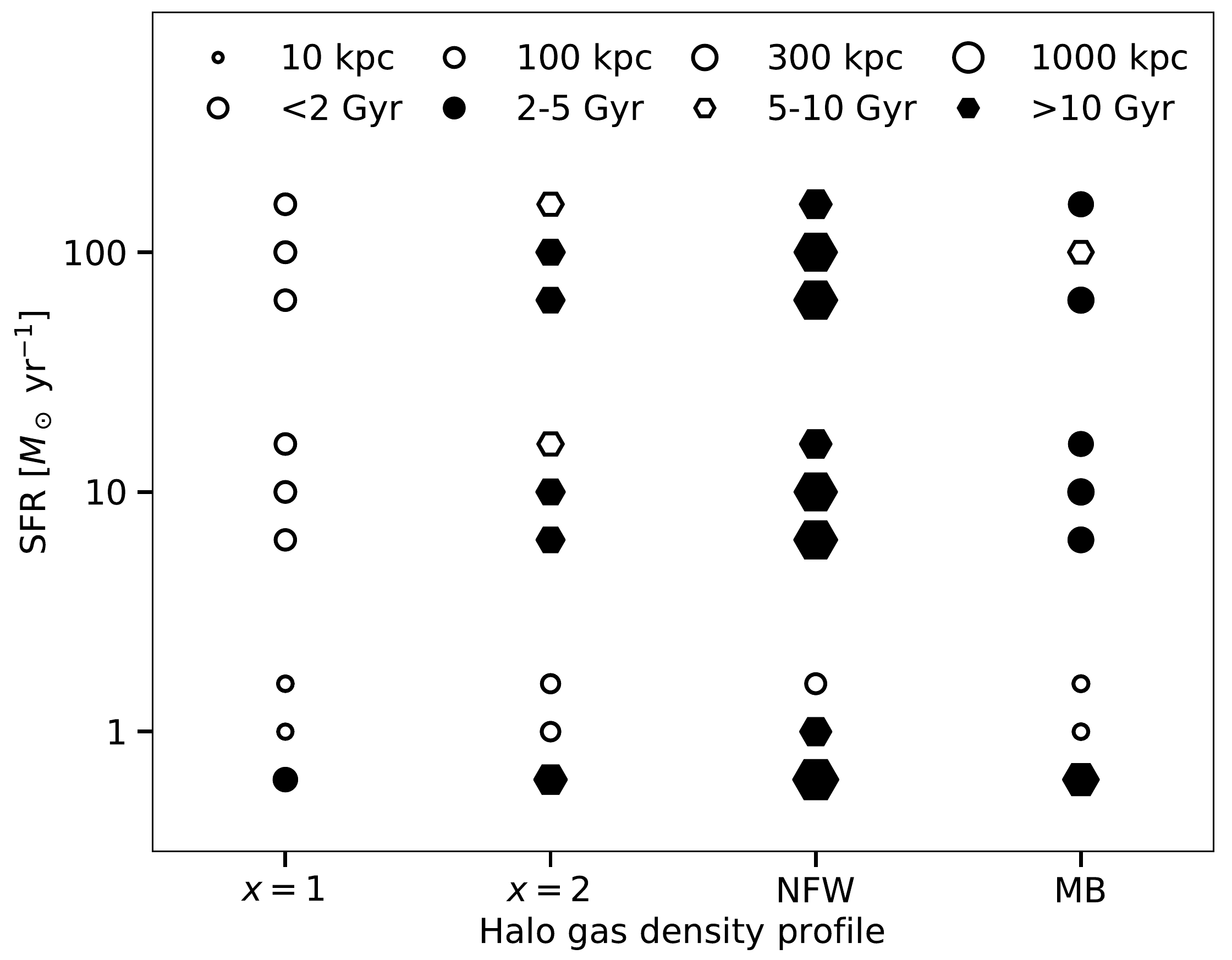}
\caption{A summary plot showing how far the modeled bubbles travel into the CGM and how long they spend past 100 kpc. Each group of three points are models run with the SFR and halo gas density profile that they align with on the vertical and horizontal axes, respectively. Each point within the group of three is for a bubble run with a different value of $\beta$, where the bottom point has $\beta=1$, the middle point has $\beta=2$, and the top point has $\beta=4$. The size of the points is logarithmically scaled to the maximum distance reached by the bubble, and the point shape and color indicate the time spent beyond 100 kpc: open circle for $<2$ Gyr, filled circle for $2-5$ Gyr, open hexagon for $5-10$ Gyr, and filled hexagon for $>10$ Gyr.}
\label{fig:summaryplot}
\end{figure}

Figure~\ref{fig:summaryplot} presents a summary of how far into the galactic halo the modeled wind bubbles travel and how much time they spend beyond 100 kpc from the host galaxy, for the parameters we explore. By comparing the size of the points within each column, we see that the halo gas density profile is the strongest determining factor of how far a bubble travels into the CGM. The bubbles expanding through the NFW halo gas density profile travel significantly further than the others, and spend $>10$ Gyr beyond 100 kpc. By comparing the points within each grouping of three, we see that the value of $\beta$ can have an effect on how long the bubble spends beyond 100 kpc, with typically lower values of $\beta$ leading to longer-lived bubbles at large radii. There is no clear monotonic trend in either distance reached or time spent beyond 100 kpc with any of the parameters we explored, indicating that the physics determining the extent and longevity of the bubble depend on how the parameters couple with each other instead of on any one parameter.

For all combinations of $\beta$, SFR $>1\ M_\odot$ yr$^{-1}$, and halo gas density profile, the bubbles radiatively cool and reach large distances $\gtrsim100$ kpc, and remain there for at least $2-10$ Gyr before falling back to the galaxy, if they fall back. These bubbles also reach large masses of $2-5\times10^{10}\ M_\odot$, made up of both swept-up halo gas and wind material launched from the galaxy, and are slow, up to a few hundred km s$^{-1}$, despite being driven by wind with velocities $v_w\sim400-1000$ km s$^{-1}$. These cases are examples of our main result that \textbf{galactic winds can drive cool, massive shells to very large distances at slow speeds, where they ``hang" for a very long time, even after star formation ceases and the galaxy is no longer actively blowing a wind.}

\begin{figure}
\centering
\includegraphics[width=\linewidth]{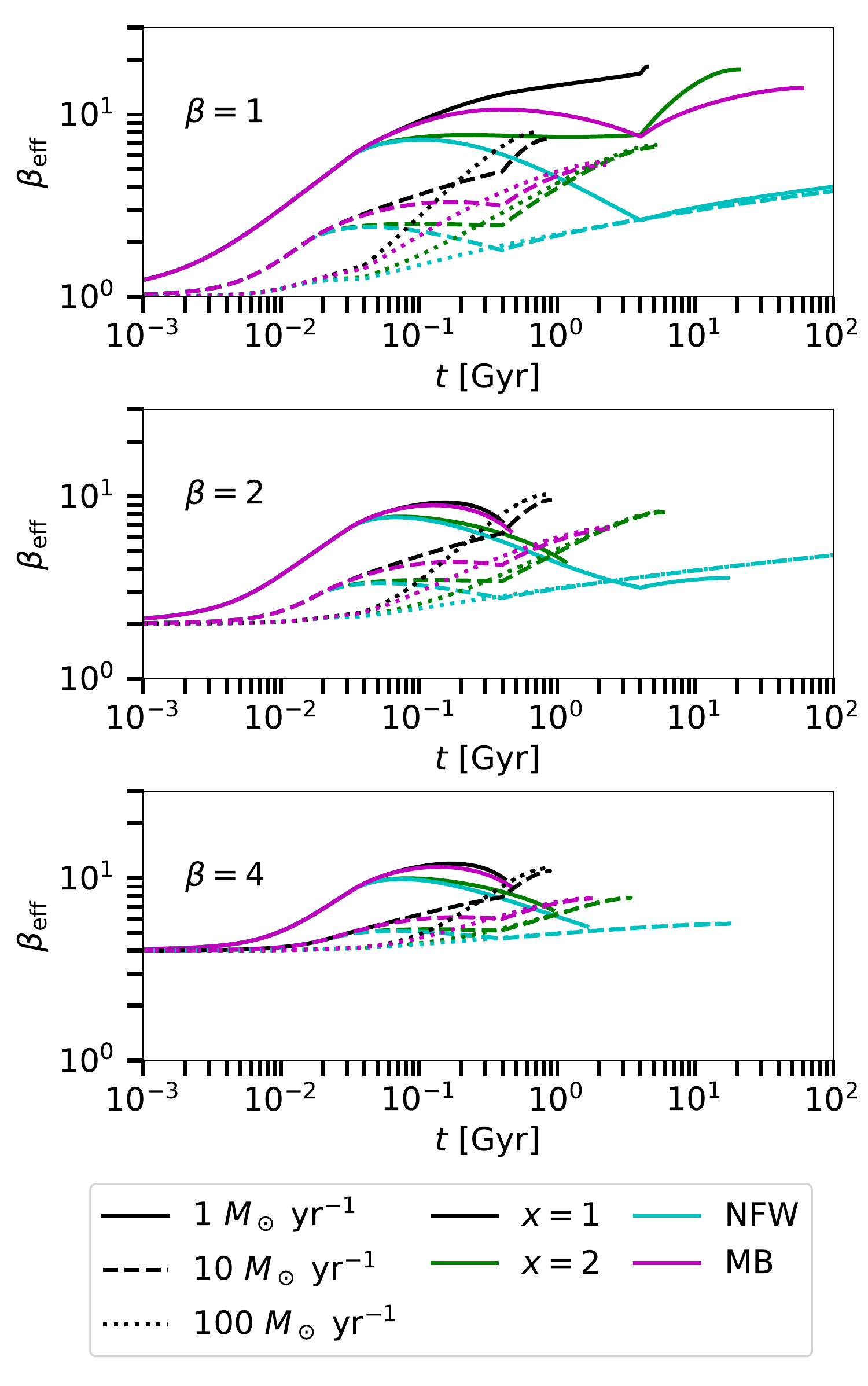}
\caption{$\beta_\mathrm{eff}=(M_w+M_\mathrm{swept})/M_\star$ as a function of time for all fiducial models. Top row shows wind models with $\beta=1$, middle row shows $\beta=2$, and bottom row shows $\beta=4$. Line color shows the halo gas density profile, see online journal for a color version of this figure. Line style indicates the SFR: $1\ M_\odot$ yr$^{-1}$ (solid), $10\ M_\odot$ yr$^{-1}$ (dashed), or $100\ M_\odot$ yr$^{-1}$ (dotted).}
\label{fig:betaeff}
\end{figure}

Note that we define $\beta$ as the mass-loading of the wind $\dot M_w/\mathrm{SFR}\sim M_w/M_\star$ at the time it is launched from the galaxy. As the bubble expands and sweeps up halo gas, the bubble mass increases drastically, thus increasing the ``effective" mass loading of the bubble above the mass loading at launch. Most bubbles have mass $M_b\sim2-5\times10^{10}\ M_\odot$, but the mass of gas turned into stars by $t_\mathrm{SFR}$ is always $4\times10^9\ M_\odot$ by design. Thus, $\beta_\mathrm{eff}\sim (M_w+M_\mathrm{swept})/M_\star\sim5-12$ for most fiducial models, where $\beta_\mathrm{eff}$ is the total mass-loading of both the wind at launch and the swept-up halo gas at the end of the integration. We plot $\beta_\mathrm{eff}$ as a function of time for all fiducial models in Figure~\ref{fig:betaeff}. The initial mass loading of the wind begins at the launch value $\beta$, but quickly increases as the bubble sweeps up halo gas mass. $\beta_\mathrm{eff}$ is not directly related to the ratio of swept-up gas to the wind gas, $M_\mathrm{swept}/M_w$, however, because most bubbles have similar $\beta_\mathrm{eff}$ by the end of integration despite being launched with different initial $\beta$. Because different fiducial models have different $\dot M_w$ (see Table~\ref{tab:models}), they have different $M_w$ and thus different $M_\mathrm{swept}/M_w$ by the end of integration. The wind mass and swept-up mass may have different metallicities from each other due to the difference in metallicity between a galaxy's interstellar medium and its gas halo; we assume $Z_\odot$ for the wind and $0.1Z_\odot$ for the halo gas. The ratio of swept-up mass to wind mass affects the overall metallicity of the bubble.

%%%%%%%%%%%%%%%%%%%%%%%%%%%%%%%%%%%%%%%%%%%%%%%%%%%%%%%%%%%%%%%%%%%%%%%%%%%%%%%%%%%%%%%%%%%%%%%%%%%
\subsection{Other Initial Conditions}
\label{sec:other_init}
%%%%%%%%%%%%%%%%%%%%%%%%%%%%%%%%%%%%%%%%%%%%%%%%%%%%%%%%%%%%%%%%%%%%%%%%%%%%%%%%%%%%%%%%%%%%%%%%%%%

We examined power-law halo gas density profiles with $x=0$, a constant density halo, and $x=3$, a very steeply-declining density profile, but do not plot these profiles in all figures because they produce too massive and not massive enough gaseous haloes, respectively. For all values of the other parameters $\beta$ and SFR, wind bubbles expanding into the constant-density gas halo ($x=0$) do not reach radii larger than $\sim70$ kpc before gravity pulls the bubble's shell back toward the galaxy. For the steeply-declining density profile with $x=3$, the bubble continues to expand indefinitely to the end of the integration time at 100 Gyr, with a large velocity $\sim400-800$ km s$^{-1}$ even at 10 Mpc from the galaxy in nearly all cases. Despite these density profiles producing unrealistic gaseous haloes, they represent effective limits on the variety of wind-blown bubble solutions and demonstrate the importance of the halo gas density profile in the evolution of the bubble. Throughout the rest of the paper, we focus on the more physically relevant halo gas density profiles shown in Figures~\ref{fig:Rv} through \ref{fig:Rv_beta4}.

We assume the thermalization efficiency $\alpha=1$, which implies that the full energy of the galaxy's supernovae is converted into heating the wind material that pressurizes and drives the bubble. Although \citet{Strickland2009} find $\alpha\sim1$ is necessary to produce the diffuse X-ray emission of M82's wind, it is interesting to consider $\alpha<1$. If only 10\% of the supernova energy heats the wind, then $\alpha=0.1$. In this case, the bubble launch velocities are lower by a factor of $\sqrt{10}\sim3$ (equation~\ref{eq:vl}), and the bubbles across all values of $\beta$, SFR, and halo gas density profile barely reach $10-40$ kpc. However, because of the low energy being supplied to the bubble by the low heating efficiency, all bubbles with $\alpha=0.1$ radiatively cool early on in their evolution. An $\alpha=0.3$ allows the bubbles to expand further than $\alpha=0.1$, so that those bubbles with $\beta=1$ and SFR $=10$ or $100\ M_\odot$ yr$^{-1}$ expand up to $\sim100-200$ kpc over $1-5$ Gyr, while still radiatively cooling early on in their evolution. However, these bubbles have lower masses $7\times10^9-10^{10}\ M_\odot$ before falling back to the galaxy. No other values of $\beta$ and SFR allow the bubbles to expand over 100 kpc when $\alpha=0.3$. We focus on $\alpha=1$ throughout this paper but caution that $\alpha$ is observationally poorly constrained and may take other values.

%%%%%%%%%%%%%%%%%%%%%%%%%%%%%%%%%%%%%%%%%%%%%%%%%%%%%%%%%%%%%%%%%%%%%%%%%%%%%%%%%%%%%%%%%%%%%%%%%%%
\subsection{Maximum $\beta$}
\label{sec:max_beta}
%%%%%%%%%%%%%%%%%%%%%%%%%%%%%%%%%%%%%%%%%%%%%%%%%%%%%%%%%%%%%%%%%%%%%%%%%%%%%%%%%%%%%%%%%%%%%%%%%%%

For a hot wind expanding adiabatically out of a galaxy, \citet{Thompson2016} showed there is a critical value of $\beta$ that places the cooling radius of the hot, pre-shocked wind, $r_\mathrm{cool}$, at the launch radius of the wind, $R_l$. Values of $\beta\gtrsim\beta_\mathrm{crit}$ do not produce a supersonic wind that drives a galactic wind bubble, but instead a subsonic ``breeze" that does not escape the galaxy \citep{Silich2011,Bustard2016}. Scaling $\beta_\mathrm{crit}$ to our fiducial parameters \citep[see equation~8 in][]{Thompson2016} gives
\begin{equation}
\beta_\mathrm{crit}\simeq 5 \alpha^{0.73}\left(\frac{R_l}{1\ \mathrm{kpc}}\right)^{0.27}\left(\frac{\mathrm{SFR}}{1\ M_\odot\ \mathrm{yr}^{-1}}\right)^{-0.27}. \label{eq:betacrit}
\end{equation}
When $\alpha=1$, $\beta_\mathrm{crit}\sim 5,\ 2.7,$ or $1.5$ for SFR $=1,\ 10,$ or $100\ M_\odot$ yr$^{-1}$, respectively. Figure~\ref{fig:Rv_beta2} shows bubbles with $\beta>\beta_\mathrm{crit}$ for SFR $=100\ M_\odot$ yr$^{-1}$ and Figure~\ref{fig:Rv_beta4} shows bubbles with $\beta>\beta_\mathrm{crit}$ for SFR $>1\ M_\odot$ yr$^{-1}$, so these bubbles may be physically unrealistic if driven purely by an adiabatic hot wind. However, we include them in our analysis so as not to rule out other types of wind driving, such as radiation pressure, cosmic rays, or non-thermal winds (see \S\ref{sec:other_feedback}). We thus stress that only those models with $\beta<\beta_\mathrm{crit}$ can be driven by adiabatic hot winds, and other driving mechanisms may be necessary to produce other galactic wind bubbles with higher mass loading.

%%%%%%%%%%%%%%%%%%%%%%%%%%%%%%%%%%%%%%%%%%%%%%%%%%%%%%%%%%%%%%%%%%%%%%%%%%%%%%%%%%%%%%%%%%%%%%%%%%%
\subsection{Reverse Shock Radius}
%%%%%%%%%%%%%%%%%%%%%%%%%%%%%%%%%%%%%%%%%%%%%%%%%%%%%%%%%%%%%%%%%%%%%%%%%%%%%%%%%%%%%%%%%%%%%%%%%%%

\begin{figure*}
\begin{minipage}{175mm}
\includegraphics[width=\linewidth]{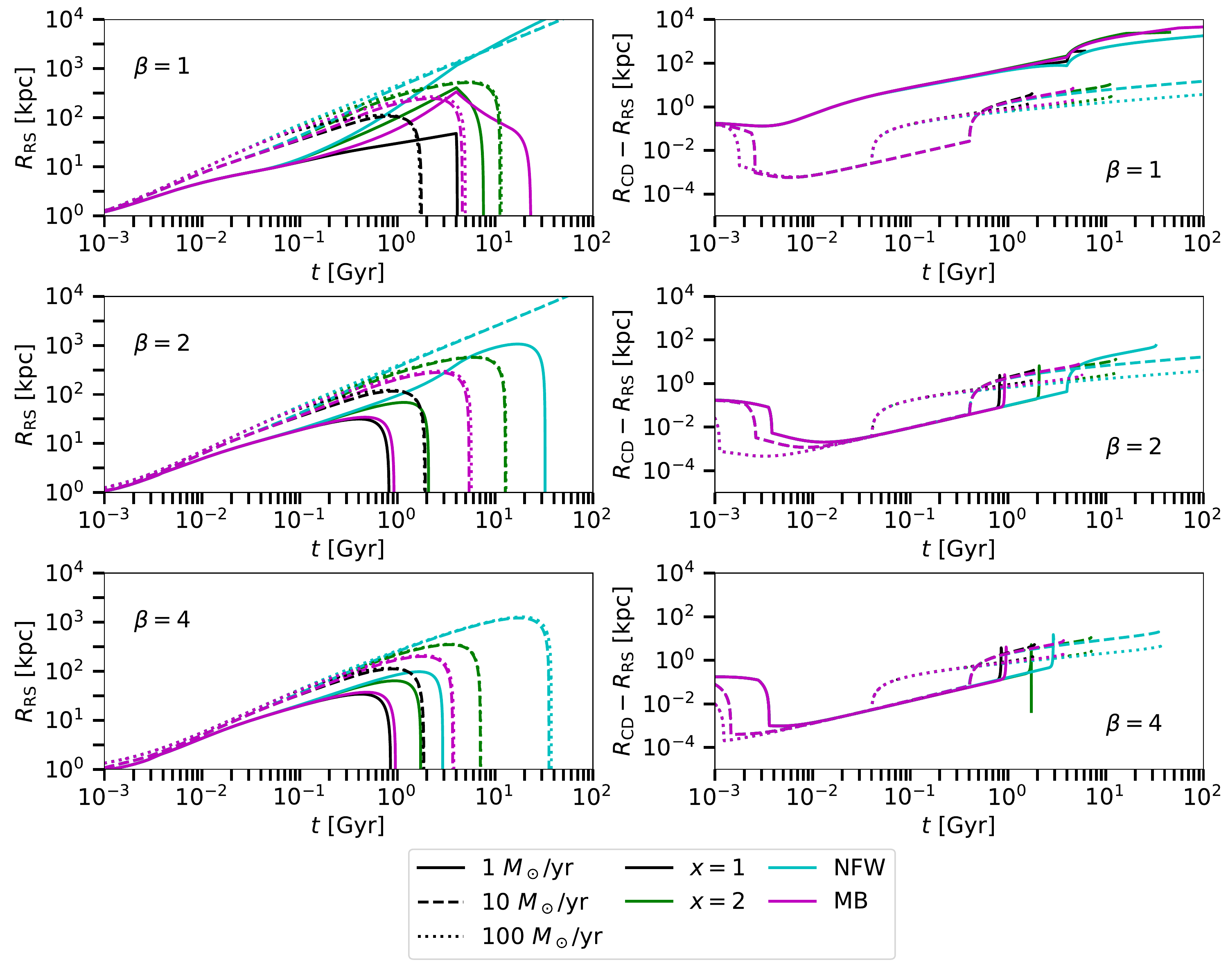}
\caption{The distance of the reverse shock from the centre of the galaxy (left panels) and the thickness of the shocked wind region (right panels) as functions of time are shown for all wind models. Top row shows wind models with $\beta=1$, middle row shows $\beta=2$, and bottom row shows $\beta=4$. Line color shows the halo gas density profile, see online journal for a color version of this figure. Line style indicates the SFR. Decreases in $R_\mathrm{CD}-R_\mathrm{RS}$ (right panels) occur when the shocked wind radiatively cools and so the distance between the reverse shock and the contact discontinuity is compressed. Sudden increases in $R_\mathrm{CD}-R_\mathrm{RS}$ occur when the star formation shuts off, and the shocked wind begins to expand adiabatically at its sound speed relative to the contact discontinuity. Very low values of $R_\mathrm{CD}-R_\mathrm{RS}$ indicate shocked wind regions that have radiated most of their thermal energy and therefore must be compressed to balance the lower thermal pressure of the shocked wind with the ram pressure of the unshocked wind.}
\label{fig:Rrsvt}
\end{minipage}
\end{figure*}

The shocked wind region between the reverse shock and the contact discontinuity radiatively cools in most of our fiducial bubbles to produce cool gas that may give rise to the low-ionization state metals observed in absorption. The reverse shock radius, $R_\mathrm{RS}$, and the thickness of the shocked wind region, $R_\mathrm{CD}-R_\mathrm{RS}$, are shown as functions of time in the left and right columns, respectively, of Figure~\ref{fig:Rrsvt}, and different values of $\beta$ are shown in each row, increasing from top to bottom. When the shocked wind radiatively cools, the shocked wind region quickly compresses to balance the wind ram pressure with the lower thermal pressure in the shocked wind region, producing the sudden decrease in $R_\mathrm{CD}-R_\mathrm{RS}$ at early times in all models other than $\beta=1$ and SFR $=1\ M_\odot$ yr$^{-1}$. Even after the shocked wind has cooled and compressed, if star formation is still ongoing, more energy and mass is being deposited into the shocked wind region, so the thickness of this region can continue to grow at a steady rate. Sharp increases in $R_\mathrm{CD}-R_\mathrm{RS}$ occur either when the bubble falls back to the galaxy or when the star formation shuts off, and the previously-shocked wind begins adiabatically expanding at its sound speed relative to the contact discontinuity, no longer constrained by the wind ram pressure. After the wind shuts off, the region internal to the reverse shock is empty of gas, so $R_\mathrm{RS}$ marks the inner radius of the previously-shocked wind as it expands, while the outer radius is still marked by $R_\mathrm{CD}$.

If cooled, the shocked wind region is $\sim1-10$ pc thick at early times $t\lesssim2$ Gyr in all models. When the wind turns off at $t=t_\mathrm{SFR}$, the shocked wind expands to $R_\mathrm{CD}-R_\mathrm{RS}\sim1-10$ kpc by the end of integration or by the time the bubble falls back to the galaxy. When $R_\mathrm{CD}-R_\mathrm{RS}$ is at its minimum, it is $\sim10^{-4}R_\mathrm{CD}$; the shell of the bubble is extremely thin compared to its radius. Even at its maximum, typically immediately before the bubble falls back to the galaxy, $R_\mathrm{CD}-R_\mathrm{RS}$ is only $\sim10\%$ the radius of the bubble in most fiducial models. The one fiducial combination of $\beta$ and SFR in which the shocked wind does not radiatively cool, $\beta=1$ and SFR $=1\ M_\odot$ yr$^{-1}$, has a much thicker shocked wind region, $\sim50\%$ of the radius of the bubble at its maximum before the wind shuts off. When $t>t_\mathrm{SFR}$, $R_\mathrm{RS}$ returns to the galaxy before $R_\mathrm{CD}$ does, so the shocked wind fills 100\% of the bubble volume. However, for $\beta=1$ and SFR $=1\ M_\odot$ yr$^{-1}$, the shocked wind does not radiatively cool and thus cannot produce the low-ionization state metal absorption lines seen in absorption that arise from lower temperature gas, unless it mixes with cool gas in the shocked halo gas region (see \S\ref{sec:instab} for Rayleigh-Taylor instability mixing).

In all cases where the bubble eventually falls back to the galaxy, the reverse shock reaches the galaxy first. If the wind has shut off and the contact discontinuity is falling back toward the galaxy, then $R_\mathrm{RS}$ will fall back even faster, because the shocked wind expands relative to the contact discontinuity. In the cases where the bubble does not fall back to the galaxy in the 100 Gyr that we integrate, such as some of the bubbles expanding into the NFW halo gas density profile, then even though the inner edge of the shocked wind expands away from the contact discontinuity at its sound speed, it cannot fall all the way back to the galaxy because the contact discontinuity is traveling faster than the shocked wind's sound speed.

%%%%%%%%%%%%%%%%%%%%%%%%%%%%%%%%%%%%%%%%%%%%%%%%%%%%%%%%%%%%%%%%%%%%%%%%%%%%%%%%%%%%%%%%%%%%%%%%%%%
\subsection{Forward Shock Radius}
%%%%%%%%%%%%%%%%%%%%%%%%%%%%%%%%%%%%%%%%%%%%%%%%%%%%%%%%%%%%%%%%%%%%%%%%%%%%%%%%%%%%%%%%%%%%%%%%%%%

The swept-up halo gas is located between the contact discontinuity and the forward shock. If the cooling time for this gas is shorter than the bubble's advection time, it can also cool and contribute to the observed low-ionization state metal line absorption. We find the forward shock radius, $R_\mathrm{FS}$, by assuming pressure balance across the contact discontinuity such that the pressure in the shocked wind region is equal to the pressure in the shocked halo gas. The temperature of the shocked halo gas, $T_\mathrm{SHG}$, is the post-shock temperature, which depends only on the velocity of the forward shock under the strong shock assumption, and we assume the forward shock velocity is equal to the contact discontinuity velocity. The temperature is given by
\begin{equation}
T_\mathrm{SHG}=\frac{3}{16}\frac{\mu m_p}{k_\mathrm{B}}v_\mathrm{CD}^2.
\end{equation}
Using the ideal gas law, the assumption that $P_\mathrm{SW}=P_\mathrm{SHG}$, and this temperature, we can find the density of the shocked halo gas\footnote{The density immediately behind the shock can also be found in the strong shock assumption as $n_\mathrm{SHG}=4 n_\mathrm{HG}$, and the density found from the ideal gas law is similar to this value. We do not use the strong shock assumption for the density as it holds only immediately behind the shock, and we are more interested in the density throughout the shocked halo gas region, which we assume to be constant and consistent with the ideal gas law.}:
\begin{equation}
n_\mathrm{SHG}=\frac{P_\mathrm{SW}}{k_\mathrm{B} T_\mathrm{SHG}}, \label{eq:nSHG}
\end{equation}
and we assume the density of the shocked halo gas is constant throughout the region. The total mass in the shocked halo gas is the swept-up mass of halo gas, so from the density and mass we can find the location of the forward shock:
\begin{equation}
R_\mathrm{FS} = \left(\frac{M_\mathrm{swept}}{\frac{4 \pi}{3} \mu m_p n_\mathrm{SHG}} + R_\mathrm{CD}^3\right)^{1/3}. \label{eq:RFS}
\end{equation}
If the shocked halo gas cools, we immediately set its temperature to $T_\mathrm{SHG}=10^4$ K. The calculation of $R_\mathrm{FS}$ then proceeds from equations~(\ref{eq:nSHG}) and~(\ref{eq:RFS}) the same way.

In all cases, the shocked halo gas radiatively cools early in the evolution of the bubble, so the thickness of the shocked halo gas, $R_\mathrm{FS}-R_\mathrm{CD}$, is small, $\sim0.1-10$ pc while the wind is still blowing. When the wind shuts off, $R_\mathrm{CD}-R_\mathrm{RS}$ increases, reducing the shocked wind pressure, so $R_\mathrm{FS}-R_\mathrm{CD}$ must increase as well to keep pressure balance across the contact discontinuity. When the bubble falls back to the galaxy, or at the end of integration if it does not fall, $R_\mathrm{FS}-R_\mathrm{CD}\sim1-10$ kpc for all fiducial bubbles, or $\lesssim1-10\%$ of the bubble radius.

Because the forward shock is close to the contact discontinuity at all times for all models, our assumption of thin-shell dynamics is valid. The very thin structure of the shocked halo gas also implies that the absorption velocities produced by cooled gas in this region would not appear different from the absorption velocities produced by the cooled shocked wind gas, as the two regions of gas are moving at approximately the same velocity and are approximately co-spatial. Our predictions of the velocity centres and widths of the absorption features are therefore not changed much by including absorption from the shocked halo gas in addition to the absorption from the shocked wind. However, the shocked halo gas region may have different column densities and metallicities (see \S\ref{sec:den_temp}). If the shocked halo gas has cooled but the shocked wind has not, such as in our models with $\beta=1$ and SFR $=1\ M_\odot$ yr$^{-1}$, then the shocked halo gas and the shocked wind will have similar kinematics but different thermal properties, possibly producing comoving absorption of low- and high-ionization state metal lines \citep{Tripp2011}.

%%%%%%%%%%%%%%%%%%%%%%%%%%%%%%%%%%%%%%%%%%%%%%%%%%%%%%%%%%%%%%%%%%%%%%%%%%%%%%%%%%%%%%%%%%%%%%%%%%%
\subsection{Cooling of Pre-shocked Wind}
%%%%%%%%%%%%%%%%%%%%%%%%%%%%%%%%%%%%%%%%%%%%%%%%%%%%%%%%%%%%%%%%%%%%%%%%%%%%%%%%%%%%%%%%%%%%%%%%%%%

\citet{Thompson2016} calculated the radiative cooling radius of an adiabatically expanding hot wind, which depends strongly on $\beta$, as well as on the launch radius of the wind and the SFR \citep{Wang1995,Efstathiou2000,Silich2003,TenorioTagle2003}. Hot wind cooling can occur even without passing through a shock that condenses the wind and raises the density. We reproduce equation~(6) from \citet{Thompson2016}, including the factor of $\mu^{2.13}$ in equation~(3) from \citet{Schneider2018}, for our spherical wind of solar metallicity here:
\begin{equation}
r_\mathrm{cool} \simeq 4\ \mathrm{kpc}\ \frac{\alpha^{2.13}}{\beta^{2.92}}\mu^{2.13}\left(\frac{R_l}{0.3\ \mathrm{kpc}}\right)^{1.79}\left(\frac{\mathrm{SFR}}{10\ M_\odot\ \mathrm{yr}^{-1}}\right)^{-0.789}.
\end{equation}
Figure~\ref{fig:cartoon} shows the location of $r_\mathrm{cool}$ relative to other regions of the bubble. For our fiducial values of $\beta$ and SFR that are allowed for purely adiabatic hot wind (\S\ref{sec:max_beta}), the cooling radius of the pre-shock wind varies between $\sim8-440$ kpc (but $r_\mathrm{cool}<80$ kpc for all models other than $\beta=1$ and SFR $=1\ M_\odot$ yr$^{-1}$). This verifies the prediction of \citet{Thompson2016} that if the free-flowing wind radiatively cools, the shocked wind does as well. Outside of $r_\mathrm{cool}$ but within the shock at $R_\mathrm{RS}$, the pre-shock wind has cooled and thus can contribute to the low-ionization state metal absorption along any sight line that passes through it.

The cooling of the pre-shock wind is only important for observations of low-ionization state metal absorption lines before the wind turns off, which occurs early in the evolution of the bubble for the higher SFR models, which have smaller $t_\mathrm{SFR}$. We are most interested in the late-time evolution of the wind bubble, because it is only at late times that it reaches large distances from the host galaxy. In addition, cooling of the pre-shock wind does not significantly affect the motion of the bubble structure, as only the kinetic luminosity of the wind drives the bubble and the post-shock temperature is set only by the bulk velocity of the wind, which is affected by gravity more than by cooling \citep{Thompson2016}. We conclude that the cooling of the pre-shock wind is only important for producing the COS-Halos observational results at large impact parameters for a select few of the models that reach large distances before $t_\mathrm{SFR}$ and have small cooling radii, and we include the cooled pre-shocked wind for those models when calculating the column densities and absorption line velocities (\S\ref{sec:obs}).

%%%%%%%%%%%%%%%%%%%%%%%%%%%%%%%%%%%%%%%%%%%%%%%%%%%%%%%%%%%%%%%%%%%%%%%%%%%%%%%%%%%%%%%%%%%%%%%%%%%
\subsection{Instabilities}
\label{sec:instab}
%%%%%%%%%%%%%%%%%%%%%%%%%%%%%%%%%%%%%%%%%%%%%%%%%%%%%%%%%%%%%%%%%%%%%%%%%%%%%%%%%%%%%%%%%%%%%%%%%%%

Spherical shock waves are susceptible to the Vishniac instability, a type of thin-shell instability \citep{Vishniac1983,Ryu1987}. A fully 3D simulation is necessary to capture the evolution of instabilities in the wind bubble, but \citet{Vishniac1989} provide analytic estimates for the timescale of instability growth and the size range of perturbations that induce the instability for an isothermal shock. Because the shocked halo gas rapidly radiatively cools in all our fiducial models, the shock is approximately isothermal despite the immediate post-shock heating very close to the shock, and the density contrast across the shock is $\gtrsim100$, much larger than the density contrast of $\sim25$ required for the shock to be unstable \citep{Vishniac1989}. The minimum timescale is given by inverting their equation~(18a) for the maximum growth rate of instabilities, which we show here using the appropriate parameters for our model:
\begin{align}
t_\mathrm{Vishniac}&=2\left(\frac{c_{s,\mathrm{HG}}^2\sigma_\mathrm{SHG}}{-P_\mathrm{SHG}\frac{\mathrm{d}v_\mathrm{FS}}{\mathrm{d}t}}\right)^{1/2} \label{eq:Vishniac_timescale} \\
&= 20\ \mathrm{Myr}\ \left(\frac{c_{s,\mathrm{HG}}}{100\ \mathrm{km\ s}^{-1}}\right)\left(\frac{\sigma_\mathrm{SHG}}{10^{-5}\ \mathrm{g\ cm}^{-2}}\right)^{1/2} \nonumber \\
&\times\left(\frac{P_\mathrm{SHG}}{10^{-14}\ \mathrm{erg\ cm}^{-3}}\right)^{-1/2}\left(\frac{\mathrm{d}v_\mathrm{FS}/\mathrm{d}t}{10^{-6}\ \mathrm{cm\ s}^{-2}}\right)^{-1/2} \label{eq:Vishniac_timescale_scaled}
\end{align}
where $c_{s,\mathrm{HG}}$ is the speed of sound in the ambient halo gas into which the forward shock expands, $P_\mathrm{SHG}$ and $\sigma_\mathrm{SHG}=\rho_\mathrm{SHG}(R_\mathrm{FS}-R_\mathrm{CD})$ are the pressure and surface density within the shocked halo gas region, and $\frac{\mathrm{d}v_\mathrm{FS}}{\mathrm{d}t}$ is the acceleration of the forward shock. The shock is unstable when $\frac{\mathrm{d}v_\mathrm{FS}}{\mathrm{d}t}<0$, when the bubble is decelerating.

Instabilities will be induced on the timescale given by equation~(\ref{eq:Vishniac_timescale}) if perturbations of a certain size are introduced to the shock front. Equation~(18b) in \citet{Vishniac1989} gives the maximum size of a perturbation that would make the shock Vishniac-unstable, which we reproduce here with the appropriate parameters for our model:
\begin{align}
\lambda_\mathrm{max}&=2\pi c_{s,\mathrm{HG}}^2\left(\frac{-P_\mathrm{SHG}\frac{\mathrm{d}v_\mathrm{FS}}{\mathrm{d}t}}{\sigma_\mathrm{SHG}}\right)^{-1/2} \label{eq:Vishniac_maxperturb} \\
&= 6.4\ \mathrm{kpc} \left(\frac{c_{s,\mathrm{HG}}}{100\ \mathrm{km\ s}^{-1}}\right)^2\left(\frac{\sigma_\mathrm{SHG}}{10^{-5}\ \mathrm{g\ cm}^{-2}}\right)^{1/2} \nonumber \\
&\times\left(\frac{P_\mathrm{SHG}}{10^{-14}\ \mathrm{erg\ cm}^{-3}}\right)^{-1/2}\left(\frac{\mathrm{d}v_\mathrm{FS}/\mathrm{d}t}{10^{-6}\ \mathrm{cm\ s}^{-2}}\right)^{-1/2} \label{eq:Vishniac_maxperturb_scaled}
\end{align}

For our fiducial models, the timescale for Vishniac instability growth increases as the bubble grows in size, so it is most unstable early in its evolution. The timescales grow from 1 Myr to 1 Gyr during the first Gyr of the bubble's evolution, for all combinations of the fiducial initial conditions. During this time, the minimum perturbation size that would induce these instabilities grows from 1 to 100 kpc. If there are denser clouds of sizes $1-100$ kpc in the ambient halo gas, they would produce instabilities in the shell of the galactic wind bubble as the outer shock wave impacts them. These instabilities would grow on timescales of a few to hundreds of Myr, potentially fragmenting the shell over this period of time. The system of shell fragments resulting from the instability will have a distribution of velocities that could increase the velocity width of absorption observed from the system.

In addition to the thin-shell Vishniac instability, the wind bubble may be Rayleigh-Taylor unstable. The Rayleigh-Taylor instability arises when a less-dense fluid accelerates into a more-dense fluid. The Rayleigh-Taylor instability may grow at the contact discontinuity depending on the acceleration of the bubble and the ratio of the gas densities on either side of the contact discontinuity. Any type of instability can fragment thin shells, leading to a ``bubble blowout" where the shocked wind contained in the bubble by the contact discontinuity can escape through the fragments of the shell \citep{HarperClark2009,Lopez2011}. Escaping pressurized wind reduces the pressure driving of the bubble and affects the overall motion of the full bubble structure. If the shell fragments only after the shocked wind has cooled, the bubble has already transitioned from pressure-driven to momentum-driven, and the loss of pressure through holes in the shell does not alter the momentum-conserving equation of motion (equation~\ref{eq:momentumconserving}). However, heavy shell fragmentation can alter the momentum-driven equation of motion because the wind may escape through holes without depositing all of its momentum. If the shell fragments after the wind has turned off at $t=t_\mathrm{SFR}$, then the equation of motion (equation~\ref{eq:momentumconserving}) depends only on gravity and the fragmentation of the shell does not alter it. In addition to fragmenting the shell, instabilities can promote mixing between the shocked wind and the shocked swept-up halo gas.

The height of Rayleigh-Taylor fingers at the contact discontinuity can be modeled in a 1D analytic form under the assumption of self-similarity and in the absence of viscosity as \citep{Dimonte2004}
\begin{equation}
\frac{\mathrm{d}h}{\mathrm{d}t}\sim\eta g_\mathrm{eff} \left(\frac{\rho_1-\rho_2}{\rho_1+\rho_2}\right) t \label{eq:RT_growth}
\end{equation}
where $h$ is the height of the instability, $g_\mathrm{eff}$ is the net acceleration felt by a particle sitting on the contact discontinuity, $\rho_1$ and $\rho_2$ are the densities on either side of the boundary defined such that $\rho_1>\rho_2$, and $\eta$ is a numerical value; we take $\eta\sim0.1$ \citep{Dimonte2004}. Rayleigh-Taylor growth occurs only when $g_\mathrm{eff}$ points from the high-density region to the low-density region, where $g_\mathrm{eff}$ is given by \citep{Pizzolato2006}
\begin{equation}
g_\mathrm{eff}=-\frac{GM_\mathrm{enc}(R_\mathrm{CD})}{R_\mathrm{CD}^2}-\frac{\mathrm{d}v_\mathrm{CD}}{\mathrm{d}t}
\end{equation}
and the sign of $g_\mathrm{eff}$ indicates whether it points outward (positive) or inward (negative). Rayleigh-Taylor instability fingers grow when the density of the shocked halo gas is greater than the density of the shocked wind, $\rho_\mathrm{SHG}>\rho_\mathrm{SW}$, and $g_\mathrm{eff}<0$. The net acceleration is negative either when the bubble is accelerating, $\frac{\mathrm{d}v_\mathrm{CD}}{\mathrm{d}t}>0$, or when the bubble is decelerating, $\frac{\mathrm{d}v_\mathrm{CD}}{\mathrm{d}t}<0$, and $\left|\frac{GM_\mathrm{enc}(R_\mathrm{CD})}{R_\mathrm{CD}^2}\right|>\left|\frac{\mathrm{d}v_\mathrm{CD}}{\mathrm{d}t}\right|$. Instabilities can also grow when $\rho_\mathrm{SHG}<\rho_\mathrm{SW}$ and $g_\mathrm{eff}>0$, which occurs when the bubble is decelerating and $\left|\frac{GM_\mathrm{enc}(R_\mathrm{CD})}{R_\mathrm{CD}^2}\right|<\left|\frac{\mathrm{d}v_\mathrm{CD}}{\mathrm{d}t}\right|$.

Because we set the location of the forward shock by requiring pressure balance between the shocked wind and shocked halo gas, the densities in the shocked wind and the shocked swept-up halo gas are equivalent when both regions of the bubble have radiatively cooled to the same temperature $T\sim10^4$ K. The shocked swept-up halo gas just outside the contact discontinuity radiatively cools almost immediately, so the densities are non-equal only before the shocked wind has radiatively cooled. For all models other than $\beta=1$ and SFR $=1\ M_\odot$ yr$^{-1}$, the shocked wind cools early, so there is not much time for instabilities to grow. During the brief time when the shocked halo gas has cooled but the shocked wind has not, the bubble is accelerating, so $g_\mathrm{eff}<0$, and the density of the shocked halo gas is always larger than the density of the shocked wind for all fiducial models, $\rho_\mathrm{SHG}>\rho_\mathrm{SW}$. This causes Rayleigh Taylor instabilities to grow, but for most models, they grow for only $\sim$few Myr at most and do not reach sizes larger than $\sim20-30$ pc, as estimated by equation~(\ref{eq:RT_growth}) before the shocked wind cools and the shell is no longer unstable.

The shell fragments if the height of the instabilities reaches the thickness of the shocked wind region \citep{Samui2008}. The thickness of the shocked wind just before it cools, when the Rayleigh-Taylor instabilities reach their maximum height, is typically $\sim100$ pc (Figure~\ref{fig:Rrsvt}). However, the shocked wind collapses when it cools and $R_\mathrm{CD}-R_\mathrm{RS}$ shrinks to $\sim5-10$ pc, smaller than the height of the instabilities. Even though the instabilities are no longer growing when the shocked wind region shrinks, the perturbations in the shell left by the Rayleigh-Taylor fingers could cause the shell to fragment. The Vishniac instability causes fragments at the same scale as the perturbations that introduce the instability, anywhere from a few to 100 kpc. The shell is certainly thinner than the scale of these perturbations after it has cooled, so the Vishniac instability can fragment the shell as well. A full 3-dimensional exploration of the galactic wind bubble structure is necessary to fully model the instability growth and fragmentation of the shell, and to follow the evolution of the fragments. We analyze the evolution of the wind bubble without instabilities and fragmentation, but caution that these processes may alter the evolution of the bubble significantly. \citet{Sarkar2015} find that instabilities in the shell cause a multiphase wind structure to form, promoting mixing with the halo gas, and some resulting clumps may fall back to the galaxy while others continue further into the halo.

%%%%%%%%%%%%%%%%%%%%%%%%%%%%%%%%%%%%%%%%%%%%%%%%%%%%%%%%%%%%%%%%%%%%%%%%%%%%%%%%%%%%%%%%%%%%%%%%%%%
\section{Comparison with Observations}
\label{sec:obs}
%%%%%%%%%%%%%%%%%%%%%%%%%%%%%%%%%%%%%%%%%%%%%%%%%%%%%%%%%%%%%%%%%%%%%%%%%%%%%%%%%%%%%%%%%%%%%%%%%%%

The COS-Halos survey \citep{Tumlinson2013} investigated the haloes of 44 galaxies using background quasar spectra observed with the Cosmic Origins Spectrograph on the \emph{Hubble Space Telescope}. The sight lines toward the quasars had impact parameters ranging from $20-154$ kpc from the foreground galaxies, and 75\% of the spectra revealed absorption lines of low or intermediate ionization state metals. \citet{Werk2014} estimated a lower limit on the total mass of cool gas of $\sim6.5\times10^{10}\ M_\odot$ in the CGM out to 300 kpc as traced by H I absorption column densities. \citet{Keeney2017} observed 16 quasar spectra probing the CGM of 37 galaxies with impact parameters ranging from $7-505$ kpc, and estimated a cool gas mass of $\sim3.2\times10^{10}\ M_\odot$ in the CGM. As evidenced by Figures~\ref{fig:Rv},~\ref{fig:Rv_beta2}, and~\ref{fig:Rv_beta4}, under certain conditions our models can produce cooled wind bubbles with masses $10^{10}-10^{11}\ M_\odot$ at several hundred kpc from galaxies, consistent with either \citet{Werk2014} or \citet{Keeney2017}. In addition, the presence of cool CGM at large impact parameters in the COS-Halos observations persists when viewing passive, non-star-forming galaxies \citep[although, there is a bimodality in higher ionization lines like O VI between star-forming and passive galaxies, see][]{Tumlinson2011}. Our model is consistent with these observations, as shells can spend an extended period of time at large distances from the host galaxy, even after star formation shuts off and the host galaxy would be observed as passive.

%%%%%%%%%%%%%%%%%%%%%%%%%%%%%%%%%%%%%%%%%%%%%%%%%%%%%%%%%%%%%%%%%%%%%%%%%%%%%%%%%%%%%%%%%%%%%%%%%%%
\subsection{Densities and Temperatures}
\label{sec:den_temp}
%%%%%%%%%%%%%%%%%%%%%%%%%%%%%%%%%%%%%%%%%%%%%%%%%%%%%%%%%%%%%%%%%%%%%%%%%%%%%%%%%%%%%%%%%%%%%%%%%%%

The COS-Halos observations and those in \citet{Keeney2017} provide column densities, which depend on the thickness of the absorbing region, its density, and its geometry. Column densities are degenerate with physical density in the absorption regions, impact parameter of the quasar spectrum probing the bubble, and the thickness of the cooled gas regions $R_\mathrm{CD}-R_\mathrm{RS}$, $R_\mathrm{FS}-R_\mathrm{CD}$, and $R_\mathrm{RS}-r_\mathrm{cool}$, where the latter is only present when $t<t_\mathrm{SFR}$. We use the modeled physical evolution of the shell to derive observables.

Modeled column densities depend on not just the physical structure of the bubble, but also on the age of the bubble when it is observed and if the wind is still blowing during observation. We eliminate the dependence on impact parameter by calculating the area-averaged column density over the full shell. At small impact parameters, the column density will be smaller than this average because impact parameters that pass close to the middle of the shell see the thinness of the shell without projection effects. Impact parameters $b$ with $R_\mathrm{RS}<b<R_\mathrm{FS}$ graze the bubble's shell and observations with these impact parameters will reveal a higher column density of cool gas. However, $R_\mathrm{FS}-R_\mathrm{RS}\sim1-10\%$ of the bubble's radius, so is it statistically unlikely for a line of sight with a random impact parameter to satisfy $R_\mathrm{RS}<b<R_\mathrm{FS}$. When $t<t_\mathrm{SFR}$, the pre-shock wind outside its cooling radius $r_\mathrm{cool}$ contributes to the total cool gas column density. The shocked wind and shocked swept-up halo gas contribute to the cool gas column density only after those regions of the bubble have cooled, which occurs at early times $t\sim1$ Myr for all models where cooling occurs at all. The models with $\beta=1$ and SFR $=1\ M_\odot$ yr$^{-1}$ are the only models where the shocked wind does not radiatively cool and does not contribute to the cool gas column density.

\begin{figure*}
\begin{minipage}{175mm}
\centering
\includegraphics[width=\linewidth]{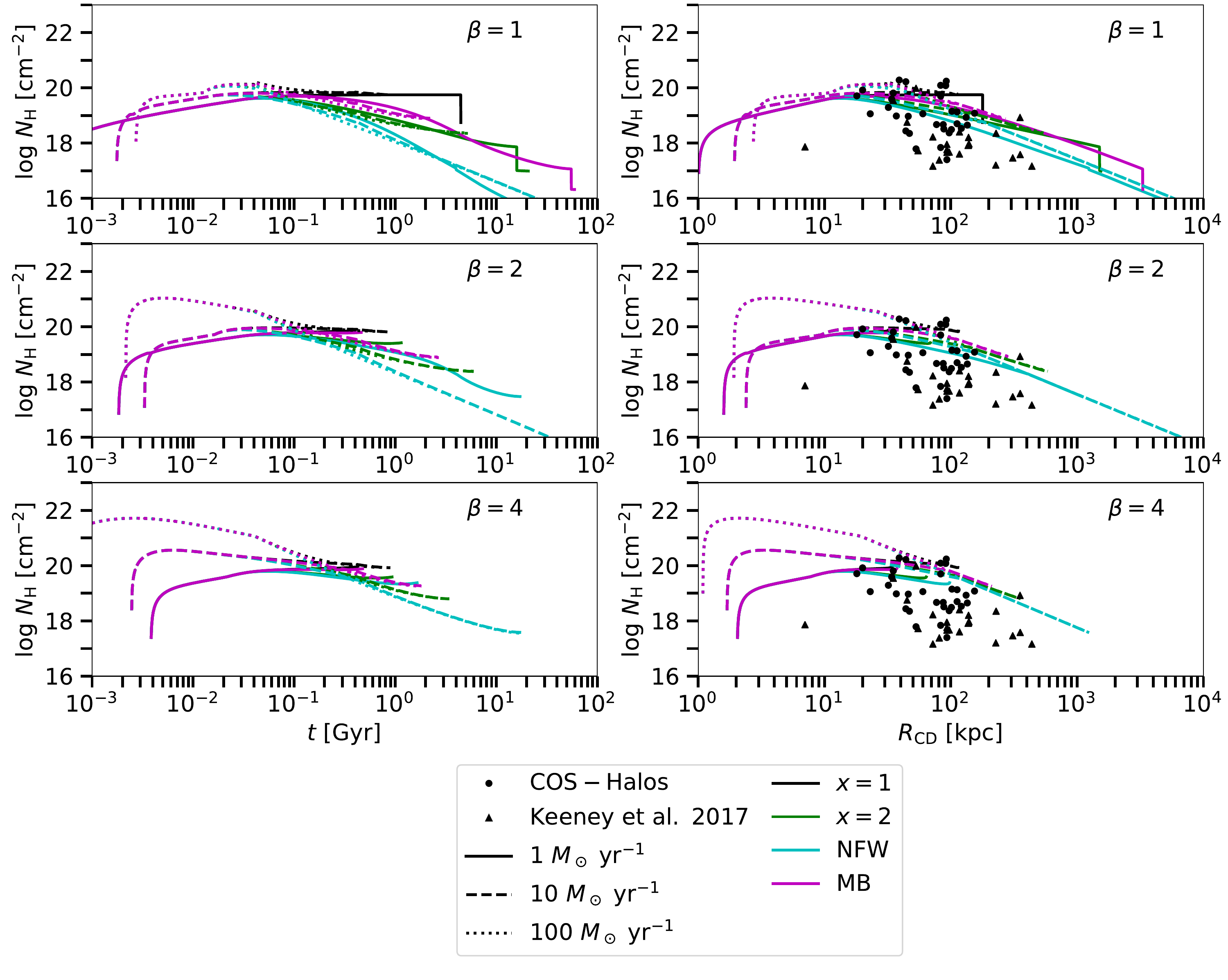}
\caption{The area-averaged column density of hydrogen combined between the shocked wind region, the shocked halo gas region, and the pre-shock cooled wind as a function of time in the left column and as a function of the location of the contact discontinuity in the right column, for models with $\beta=1$ (top row), $\beta=2$ (middle row), and $\beta=4$ (bottom row). Curve colors and styles are as in Figure~\ref{fig:Rrsvt}. Black circular points indicate the observed ionization-corrected hydrogen column densities in the COS-Halos sample and black triangle points indicate observed ionization-corrected columns in the \citet{Keeney2017} sample, as functions of impact parameter. Model curves are ``right-limits" --- a model can predict the observations if it lies to the right of the points.}
\label{fig:NHvtR}
\end{minipage}
\end{figure*}

Figure~\ref{fig:NHvtR} shows the area-averaged cool gas column density of hydrogen with $\beta=1$ (top row), $\beta=2$ (middle row), and $\beta=4$ (bottom row), as functions of time (left panels) and functions of $R_\mathrm{CD}$ (right panels). Note that the right panels do not show column density as a function of impact parameter from the host galaxy in the model curves; they show column density as a function of the location of the contact discontinuity, which changes with time such that larger $R_\mathrm{CD}$ generally corresponds to later times. When the shell is at a distance $R_\mathrm{CD}$ as shown on the horizontal axis, the observed column density will be approximately that on the vertical axis no matter the impact parameter of the observation, as long as $b<R_\mathrm{FS}$. In this way, the model curves in the right panels can be viewed as ``right-limits" --- the observed column will, on average, be the plotted $\log N_\mathrm{H}$ for any impact parameter that falls to the left of the model curve.

The column density increases as both the wind mass and swept-up halo gas mass increase, then begins to decrease slowly as the bubble expands further and the bubble's mass is spread thin, despite the always-increasing total mass of the bubble. When the bubble passes the cooling radius of the pre-shock wind at $r_\mathrm{cool}$, the mass of cool pre-shock wind is added to the column density, and there is a slight increase in the area-averaged column density. The location of the increase in $\log N_\mathrm{H}$ in the right panels corresponds to the shell of the bubble passing $r_\mathrm{cool}$, if it reaches $r_\mathrm{cool}$ while $t<t_\mathrm{SFR}$. This increase is visible in the $\beta=1$ and SFR $=100\ M_\odot$ yr$^{-1}$ (dotted) curves at 10 kpc, for example. For many combinations of $\beta$ and SFR, the bubble either does not reach $r_\mathrm{cool}$ before the wind shuts off, or $r_\mathrm{cool}$ is located at (or below, see \S\ref{sec:max_beta}) the wind launch radius.

The black points in the right panels show the total hydrogen column density reported by \citet{Werk2014} and \citet{Keeney2017}. The data are shown as functions of impact parameter, unlike the model curves, which are presented as functions of $R_\mathrm{CD}$. Because the model curves fall to the right of the majority of data points, they predict the column density observations well, but tend to somewhat over-predict $\log N_\mathrm{H}$ for $\beta=4$. \citet{Prochaska2017} give stricter estimates on the H I column density for a handful of the COS-Halos galaxies shown in Figure~\ref{fig:NHvtR} that generally revise the column densities up by factors of $\sim0.2-0.5$ dex, so our models may not as strongly over-predict $\log N_\mathrm{H}$ when compared to the revised values. The models can reproduce the smallest observed values of $\log N_\mathrm{H}$ if these columns are due to observing a very large bubble, as the largest bubbles have the smallest column densities.

The mean redshift of the galaxies in the COS-Halos and \citet{Keeney2017} samples are $z\sim0.2$ and $z\sim0.02$, respectively, corresponding to cosmic ages $\sim11$ Gyr and $\sim13$ Gyr for a flat $\Lambda$CDM universe, so the wind-blown bubble could have had at least 10 Gyr to evolve by the time of observation. We can also compare our hydrogen column density results to measurements of circumgalactic hydrogen at higher redshift $z\sim2.3$, when the universe was only $\sim3$ Gyr old and any wind bubbles around galaxies at this redshift could not have been evolving for more than a couple Gyr. \citet{Rudie2012} measured the H I column densities of absorbers within $300$ comoving kpc of galaxies at a mean redshift of $z\sim2.3$ and found a large spread in $\log N_\mathrm{HI}\sim14-18$ at all impact parameters $50\lesssim b\lesssim300$ comoving kpc. They do not convert their measurement of $\log N_\mathrm{HI}$ to $\log N_\mathrm{H}$ so we cannot directly compare to our models, but COS-Halos \citep{Tumlinson2013} and \citet{Keeney2017} both find $\log N_\mathrm{HI}\sim13-18$ for impact parameters $b\sim50-500$ kpc, and up to $\log N_\mathrm{HI}\sim20$ for $b\lesssim50$ kpc, in rough agreement with $\log N_\mathrm{HI}$ found by \citet{Rudie2012}. The ionizing photon rate is roughly 10 times larger at $z\sim2$ than at $z=0.2$ \citep{Kollmeier2014}, so values of $\log N_\mathrm{HI}\sim14-18$ would convert to larger values of $\log N_\mathrm{H}$ at $z\sim2$ than at $z=0.2$. Our models tend to somewhat over-predict $\log N_\mathrm{H}$, so even larger values than those measured by COS-Halos and \citet{Keeney2017} could still be consistent with our models, but high-redshift wind bubbles have had significantly less time to evolve. The models with $\beta=1$ or $\beta=2$, SFR $\geq10\ M_\odot$ yr$^{-1}$, and power-law with $x=2$, NFW, or MB halo gas density profiles are the only bubbles that reach up to several hundred kpc before 1 Gyr. The targeted high-redshift galaxies are compact and have large SFR $\sim30\ M_\odot$ yr$^{-1}$, so our models with SFR $\geq 10\ M_\odot$ yr$^{-1}$ are consistent and could predict wind bubbles in these high-redshift galaxies.

\begin{figure*}
\begin{minipage}{175mm}
\centering
\includegraphics[width=\linewidth]{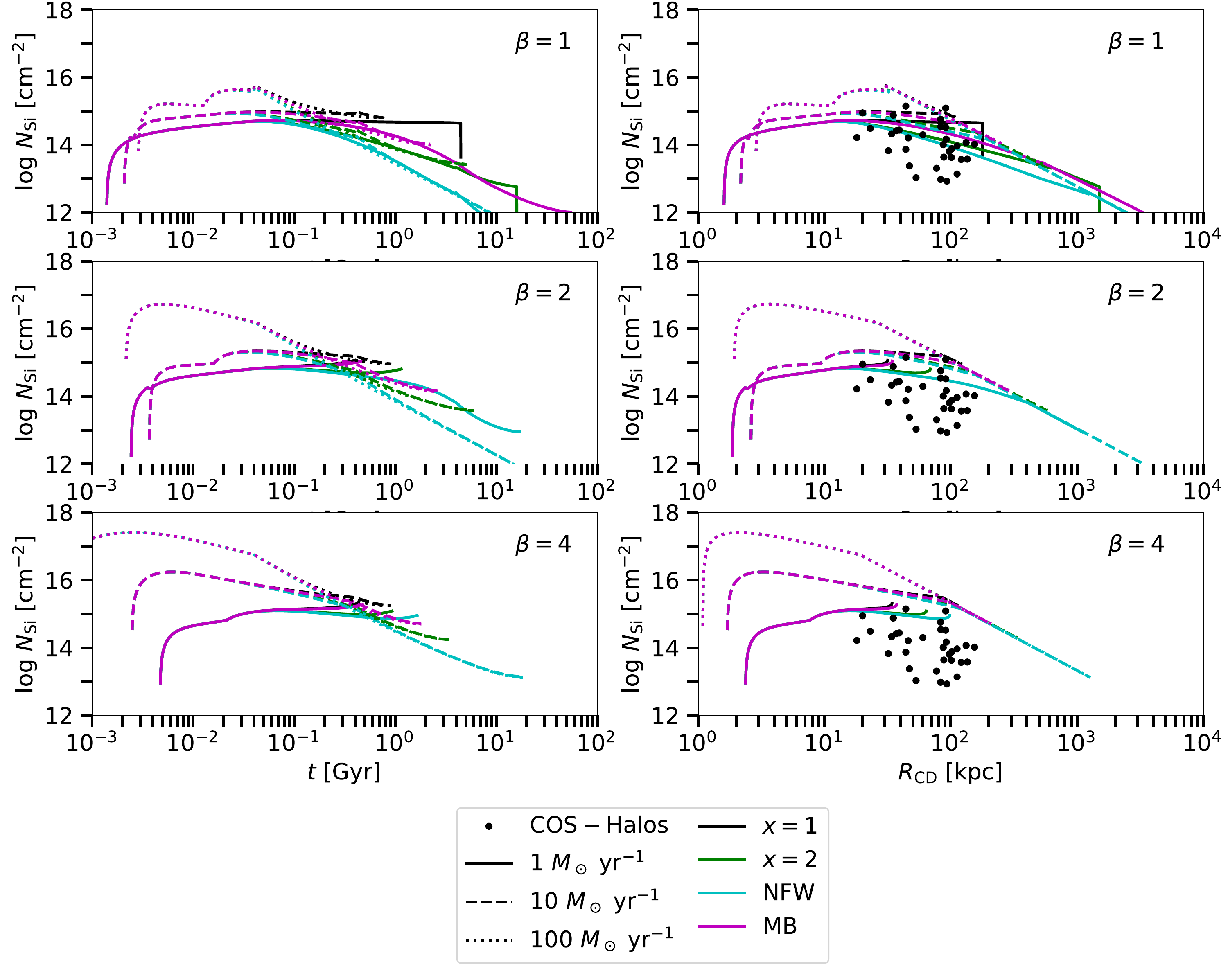}
\caption{The silicon column density (all ionization states) in the combined shocked wind region, shocked halo gas region, and pre-shock cooled wind in our models with $\beta=1$ (top row), $\beta=2$ (middle row), or $\beta=4$ (bottom row) as a function of time (left column) and as a function of the location of the contact discontinuity (right column) for model curves of as a function of impact parameter for data points. Curve colors and styles are as in Figure~\ref{fig:NHvtR}. Black points indicate the observed ionization-corrected silicon column densities in the COS-Halos sample, as a function of impact parameter. Model curves are ``right-limits" --- a model can predict the observations if it lies to the right of the points.}
\label{fig:NSivtR}
\end{minipage}
\end{figure*}

To compare our models to observed low-ionization state metal line absorption, we take silicon as an example metal. We calculate the silicon column density predicted by our models by assuming a solar abundance of Si in the wind, and a $0.1$ solar abundance in the swept-up material. The ratio of $\log N_\mathrm{Si}$ to $\log N_\mathrm{H}$ is thus mass-weighted by the proportion of the bubble that is wind material or swept-up material, with different metallicities for each. At $\sim10^4-10^{4.5}$ K, Si II and Si III are the dominant silicon ionization states, but the ionization fractions of each are dependent on whether the gas is in collisional ionization equilibrium or photoionization equilibrium \citep{Oppenheimer2013}. We forego making a choice, and compare our models' predicted $\log N_\mathrm{Si}$ to ionization-corrected $\log N_\mathrm{Si}$ reported by \citet{Werk2014}. Figure~\ref{fig:NSivtR} shows the combined $\log N_\mathrm{Si}$ of the shocked wind, shocked swept-up halo gas, and pre-shock cooled wind for $\beta=1$ (top row), $\beta=2$ (middle row), and $\beta=4$ (bottom row). The left panels show the evolution with time and the right panels show modeled $\log N_\mathrm{Si}$ as a function of $R_\mathrm{CD}$, as well as the COS-Halos $\log N_\mathrm{Si}$ as a function of impact parameter as circular points. As in Figure~\ref{fig:NHvtR}, there is a slight increase in $\log N_\mathrm{Si}$ when the bubble passes $r_\mathrm{cool}$, due to the cooled pre-shocked wind. Again, the model curves in the right panels are ``right-limits," such that any model that lies to the right of the data is consistent with the observations. In general, our models are consistent with the COS-Halos derived columns, but tend to over-predict $\log N_\mathrm{Si}$, particularly for larger values of the mass-loading $\beta$. Like for $\log N_\mathrm{H}$, our models are consistent with the smallest observed values of the column density if these columns come from observing a large bubble.

The hydrogen column densities are dominated by the swept-up shocked halo gas for bubbles blown by SFR $=1\ M_\odot$ yr$^{-1}$, but have roughly equal contributions from the shocked wind and swept-up shocked halo gas for larger SFRs. The shocked wind and swept-up shocked halo gas contribute approximately equally to the silicon column densities when SFR $=1\ M_\odot$ yr$^{-1}$, but the higher wind metallicity causes $\log N_\mathrm{Si}$ to be strongly dominated by either the pre-shock wind (when $t<t_\mathrm{SFR}$) or the shocked wind (when $t>t_\mathrm{SFR}$) for larger SFRs.

Deriving $\log N_\mathrm{H}$ and $\log N_\mathrm{Si}$ from observations requires making assumptions about the ionization state of the CGM to convert from the observed values of $\log N_\mathrm{HI}$ and, e.g., $\log N_\mathrm{SiII}$ or $\log N_\mathrm{SiIII}$. The ionization state of the gas depends on the ionizing radiation field, temperature, and density, including substructures such as dense clouds. Our smooth, spherically symmetric model does not include clouds in the CGM, so we do not compute the column density of different ionization states within cold clumps for direct comparison with observations. We caution that the comparisons between our models and the observations in Figures~\ref{fig:NHvtR} and~\ref{fig:NSivtR} depend on the assumptions and ionization corrections of \citet{Werk2014} and \citet{Keeney2017}, which may not be consistent with our model. However, it is encouraging that we find similar column densities to the observations after such corrections are made.

Finally, we check the temperature of the shocked wind. Before cooling, the temperature is derived from the pressure and density by assuming the gas is ideal. The temperature of the shocked wind is initially $\sim3\times10^7$ K for $\beta=1$, $\sim1.3\times10^7$ K for $\beta=2$, and $\sim7\times10^6$ K for $\beta=4$, and decreases only slightly as the bubble expands in all cases. Increasing $\beta$ leads to lower initial shocked wind temperatures because higher mass loading leads to slower launch speeds for the wind, and therefore slower reverse shocks and lower shocked temperatures. There is only a drastic change in the temperature if the shocked wind radiatively cools; we set the shocked wind temperature to $10^4$ K immediately when the cooling time becomes shorter than the advection time.

Although we do not perform any detailed ionization modeling, there are some interesting implications of a multiphase bubble structure. Absorption by O VI can arise from radially cooling wind \citep{Wakker2012,Bordoloi2017}, which in our model causes the shocked, cooling wind region to contract from a thick, velocity-extended region into a significantly thinner, denser region. This could potentially explain the wider velocity spread of high ionization state lines like O VI and N V that trace hotter gas seen by \citet{Muzahid2015} while maintaining the narrower velocity spread of the low ionization state lines that represent the cooled gas from the compressed region. The metallicity of low ionization absorbing clouds can indicate their origin: high metallicity cool clouds \citep{Muzahid2018} could arise from the instability-promoted fragmentation of the shocked and cooled wind, while low metallicity cool clouds could represent the shocked and cooled swept-up ambient halo gas.

%%%%%%%%%%%%%%%%%%%%%%%%%%%%%%%%%%%%%%%%%%%%%%%%%%%%%%%%%%%%%%%%%%%%%%%%%%%%%%%%%%%%%%%%%%%%%%%%%%%
\subsection{Absorption Lines}
\label{sec:lines}
%%%%%%%%%%%%%%%%%%%%%%%%%%%%%%%%%%%%%%%%%%%%%%%%%%%%%%%%%%%%%%%%%%%%%%%%%%%%%%%%%%%%%%%%%%%%%%%%%%%

We estimate the velocity centres and widths of absorption lines arising from the radiatively cooled shocked wind region, the cooled shocked swept-up halo gas, and the cooled pre-shocked wind (for $t<t_\mathrm{SFR}$), and compare to the COS-Halos absorption line data. We do not model ionization states or line profiles and thus are not concerned with the depth or shape of the absorption lines, only their velocity centre and width. We assume low-ionization state metal line absorption arises only from the cool gas.

We calculate the velocity width of the absorption lines observed at impact parameter $b$ as the difference between the fastest and slowest line-of-sight components of cool gas velocity:
\begin{equation}
v_\mathrm{width} = v_\mathrm{fast}\sqrt{\frac{R_\mathrm{fast}^2-b^2}{R_\mathrm{fast}^2}}
- v_\mathrm{slow}\sqrt{\frac{R_\mathrm{slow}^2-b^2}{R_\mathrm{slow}^2}} \label{eq:velwidth}
\end{equation}
where $v_\mathrm{fast}$ is the fastest physical velocity of the cooled gas, $R_\mathrm{fast}$ is the radius at which the cooled gas reaches the fastest velocity, $v_\mathrm{slow}$ is the slowest physical velocity of cooled gas, and $R_\mathrm{slow}$ is the radius at which the cooled gas is slowest. For example, if the only cool gas is the shocked wind and shocked halo gas, and the reverse shock is traveling at a slower velocity than the contact discontinuity (we assume the forward shock and contact discontinuity have the same velocity), then $v_\mathrm{fast}=v_\mathrm{CD}$, $R_\mathrm{fast}=R_\mathrm{FS}$, $v_\mathrm{slow}=v_\mathrm{RS}$, and $R_\mathrm{slow}=R_\mathrm{RS}$. In this way, $v_\mathrm{width}$ indicates the widest spread of velocities that could be observed in absorption by the cool gas in any region of the bubble. Note this is neither a full-width at half-maximum nor an equivalent width.

Assuming the absorption lines are symmetric, we then calculate the velocity centre of the absorption line as halfway between the fastest and slowest velocity components:
\begin{equation}
v_\mathrm{cen} = \frac{1}{2}\left[v_\mathrm{fast}\sqrt{\frac{R_\mathrm{fast}^2-b^2}{R_\mathrm{fast}^2}}
+ v_\mathrm{slow}\sqrt{\frac{R_\mathrm{slow}^2-b^2}{R_\mathrm{slow}^2}}\right]. \label{eq:vcen}
\end{equation}

\begin{figure*}
\begin{minipage}{175mm}
\centering
\includegraphics[width=\linewidth]{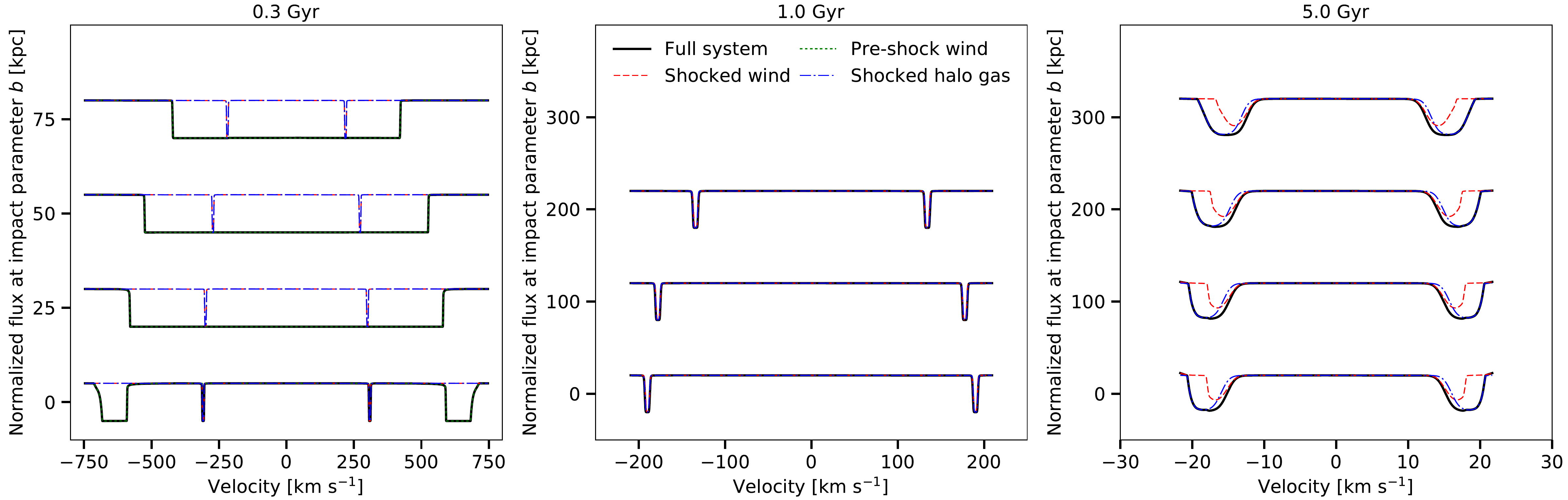}
\caption{Example normalized flux profiles of low-ionization state metal line absorption at varying impact parameters, for the particular case of $\beta=2$, SFR $=10\ M_\odot$ yr$^{-1}$, and a power-law halo gas density profile with $x=2$, at the three snapshots in time of 0.3 Gyr, 1 Gyr, and 5 Gyr (panels from left to right). Flux profiles for absorption observed at several impact parameters $b=0,\ 25,\ 50,\ 75$ kpc in the left panel and $b=0,\ 100,\ 200,\ 300$ kpc in the centre and right panels are shown, as indicated on the vertical axes. Only positive impact parameters are shown, but the symmetric structure ensures the same absorption features are present at the opposite impact parameters as well. Red dashed curves show absorption due to the shocked and cooled wind between $R_\mathrm{RS}$ and $R_\mathrm{CD}$, blue dot-dashed curves show absorption due to the shocked and cooled swept-up halo gas between $R_\mathrm{CD}$ and $R_\mathrm{FS}$, and green dotted curves show absorption due to pre-shock cooled wind between $r_\mathrm{cool}$ and $R_\mathrm{RS}$, which is only present before the wind turns off at $t=t_\mathrm{SFR}$. Solid black curves show the sum absorption from all regions of the bubble structure, which is what would be observed.}
\label{fig:example_lines}
\end{minipage}
\end{figure*}

Figure~\ref{fig:example_lines} shows an example of the normalized absorption line flux profiles as functions of line-of-sight velocity relative to the galaxy systemic velocity, for a low-ionization state metal absorption line that arises from cold gas, such as Si II or Si III. The absorption profiles are calculated assuming that the gas velocity distribution is flat between the slowest and fastest velocity within each region. Because the velocity distribution is not Gaussian, the absorption profiles do not appear Gaussian. This calculation is performed as a rough example only to provide a visualization of the bubble structure in velocity space, and represents the range of velocities over which absorption can be expected, not a physical absorption line profile. Each curve shows the flux profile, normalized between 0 and 1, vertically located at a handful of impact parameters as marked on the vertical axis. Each panel indicates a different age for the bubble at the time of observation, increasing from left to right. We use the silicon column densities (Figure~\ref{fig:NSivtR}) as representative of low-ionization metals for each region of the bubble structure. The absorption arising from different regions of the bubble are shown as different curves: red dashed curve for the shocked and cooled wind, green dotted curve for the cooled pre-shock wind (only present before wind turns off), and blue dash-dotted curve for the shocked and cooled swept-up halo gas. The black solid curve shows the sum of cool gas in all regions of the bubble and represents the general regions in velocity space where cool gas would be observed. This particular example is a bubble with $\beta=2$, SFR $=10\ M_\odot$ yr$^{-1}$, and a power-law halo gas density profile with $x=2$ (green dashed curves on other plots, for comparison), at $0.3$ Gyr (left panel), 1 Gyr (middle panel), and 5 Gyr (right panel).

In this particular case, the shocked wind and shocked swept-up halo gas both radiatively cooled very early in the bubble's evolution, before $0.3$ Gyr. The pre-shock wind cools at a radius of $r_\mathrm{cool}\sim9.3$ kpc and contributes to the absorption lines in the leftmost panel at 0.3 Gyr, which is the only panel shown here with $t<t_\mathrm{SFR}$, when the wind is still blowing. Before the wind turns off at $t_\mathrm{SFR}=0.4$ Gyr, the absorption lines with $b>r_\mathrm{cool}$ are hundreds of km s$^{-1}$ wide because these lines of sight pass entirely through cool, absorbing gas, without impacting the inner hot wind region at $r<r_\mathrm{cool}$. After the wind turns off (middle and right panels), the absorption lines become very narrow because $v_\mathrm{RS}\approx v_\mathrm{CD} \approx v_\mathrm{FS}$ and the shell is thin so few impact parameters pass entirely through cool gas without impacting the central cavity of the bubble.

The line of sight with $b=0$ kpc at $0.3$ Gyr (left panel) shows absorption with a multi-component structure due to the difference in velocity between the pre-shock cooled wind and the cooled shocked wind and halo gas that make up the shell of the bubble. At larger impact parameters $b>r_\mathrm{cool}$, the multi-component structure is wiped out because the line of sight passes entirely through cool gas so absorption is observed at all line of sight velocities. Because the velocities of the different bubble regions overlap, it is impossible to distinguish the components of absorption that represent the shell of the bubble from the components that represent the interior cooled pre-shock wind.

The line of sight with impact parameter $b=0$ kpc probes the physical velocity of the shocked wind gas directly without projection effects. At 1 Gyr (middle panel), the bubble is still expanding rapidly and absorption is observed at $v_\mathrm{CD}\sim200$ km s$^{-1}$. At larger impact parameters, projection effects produce observations of slower gas, as the shell is expanding mostly perpendicular to the line of sight. The impact parameter $b=300$ kpc does not intersect the shell because the radius of the shell at 1 Gyr is $\sim250$ kpc, so no absorption is detected at any impact parameter $b>250$ kpc. At 5 Gyr (right panel), the shell has slowed down considerably, to $\sim15$ km s$^{-1}$.

\begin{figure*}
\begin{minipage}{175mm}
\centering
\includegraphics[width=\linewidth]{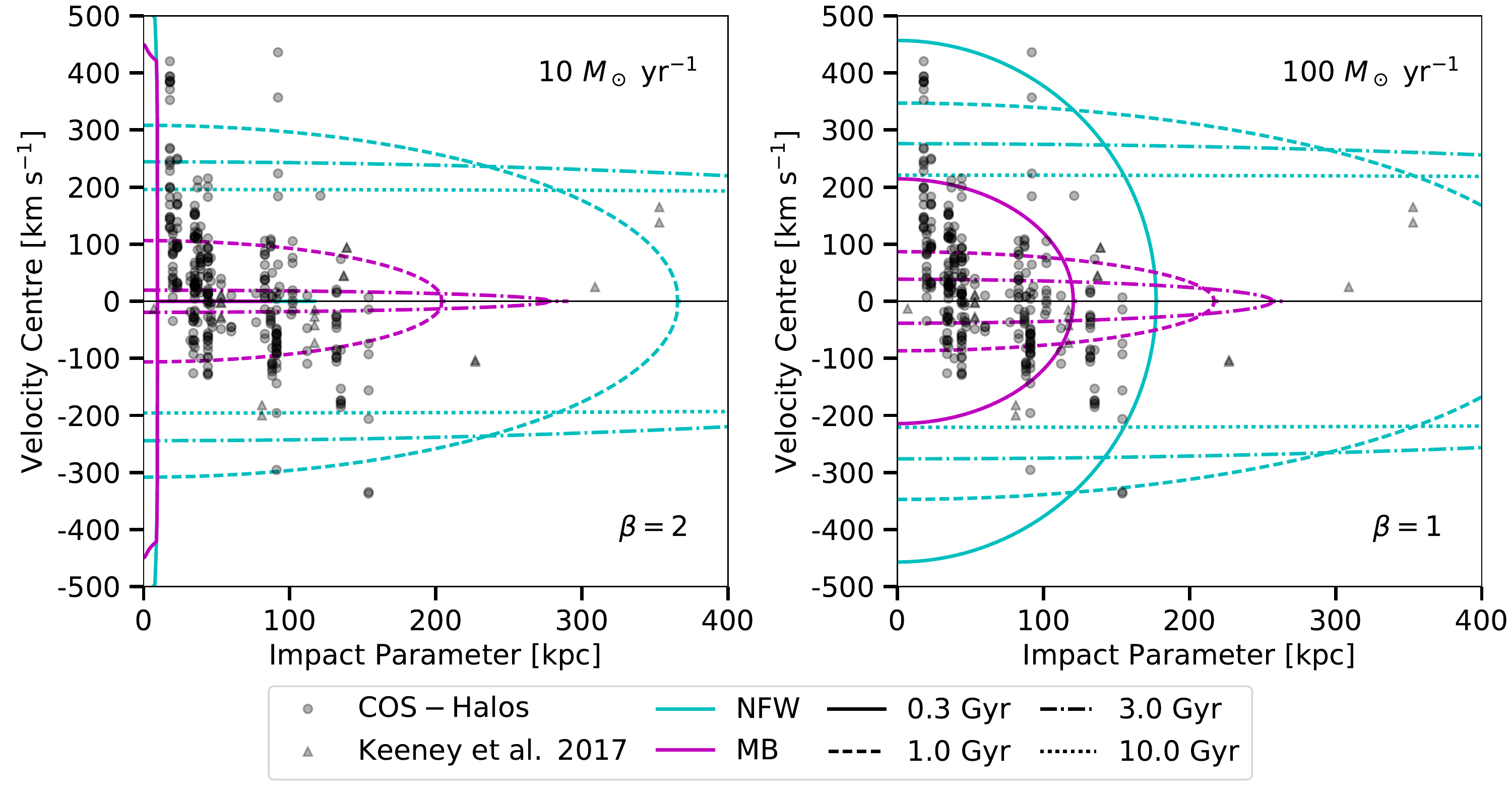}
\caption{The velocity centres of absorption lines predicted by the models as a function of impact parameter from the host galaxy compared to the velocity centres (relative to host galaxy velocity) of each component in all absorption systems in COS-Halos (translucent black circles) and \citet{Keeney2017} (translucent black triangles). Curve colors indicate models with different halo gas density profiles: MB profile (magenta, small velocities) and NFW profile (cyan, large velocities); other halo gas density profiles are omitted for readability. Curve line style indicates age of the bubble: $0.3$ Gyr (solid), 1 Gyr (dashed), 3 Gyr (dot-dashed), and 10 Gyr (dotted). Left panel shows models with SFR $=10\ M_\odot$ yr$^{-1}$ and $\beta=2$ and right panel shows models with SFR $=100\ M_\odot$ yr$^{-1}$ and $\beta=1$.}
\label{fig:vcen}
\end{minipage}
\end{figure*}

We calculate the line velocity centres and widths for the models with $\beta=1$ and SFR $=100\ M_\odot$ yr$^{-1}$ or $\beta=2$ and SFR $=10\ M_\odot$ yr$^{-1}$, as these are the models with the largest mass loadings for a hot wind (see \S\ref{sec:max_beta}) that reach large distances and have radiatively cooling shocked wind. Figure~\ref{fig:vcen} shows the velocity centre of absorption lines at four bubble ages: $0.3$ (solid), $1$ (dashed), $3$ (dot-dashed), and $10$ (dotted) Gyr, compared to the velocity centres of each absorption component in all low-ionization state metal absorption systems in COS-Halos (translucent black points) and \citet{Keeney2017} (translucent black triangles). For readability, only the MB and NFW profiles are plotted. Curves for bubbles expanding through the power law halo gas density profile with power $x=2$ are generally located between the MB and NFW curves, while the model curves for bubbles expanding through the power law halo gas density profile with $x=1$ are located at both smaller velocities and smaller impact parameters than the MB curves. The 10 Gyr curve for the MB bubble is not shown because this bubble falls back to the galaxy before 10 Gyr.

In general, the spherically symmetric structure of the bubble produces two absorption components with velocity centres symmetrically distributed around 0 km s$^{-1}$. The youngest bubbles produces absorption at larger velocities and smaller impact parameters, while only the oldest bubbles can be observed at large impact parameters, as it takes $\sim$few Gyr for the bubble to reach distances of several hundred kpc from the galaxy. A small observed absorption line velocity can be due to either a line of sight impacting the very edge of a young bubble, where the low observed velocity arises from projection effects, or a line of sight impacting any part of a physically slow bubble, where the low observed velocity is physical. Note, however, that a random distribution of impact parameters across the projected area of a bubble is unlikely to impact the very edge of the shell, so a small line velocity is more likely to arise from a physically slow, and therefore likely old, bubble. Large observed absorption line velocities must come from physically fast bubbles, which are typically younger and thus can only be observed at small impact parameters.

For SFR $=10\ M_\odot$ yr$^{-1}$, the youngest bubble age plotted is before the wind shuts off at $0.4$ Gyr, so the pre-shock cooled wind contributes to the absorption. The cooling radius $r_\mathrm{cool}\sim4.5$ kpc, so impact parameters $b>4.5$ kpc show an absorption velocity centre of 0 km s$^{-1}$, because lines of sight with these impact parameters do not impact the central cavity of the bubble and thus observe cool gas at all velocities $v<v_\mathrm{CD}$ along the line of sight. It is clear that cooled pre-shock wind alone cannot produce absorption lines consistent with the velocity of the observations. The right panels show absorption from only the cooled shocked wind and swept-up halo gas because $t_\mathrm{SFR}=0.04$ Gyr when SFR $=100\ M_\odot$ yr$^{-1}$, so even for the youngest bubble plotted at $0.3$ Gyr, the wind has shut off so there is no pre-shock wind.

The spread of predicted absorption line velocities across the various models is consistent with the spread of observed line velocities from COS-Halos and \citet{Keeney2017}. Only the youngest bubbles at $0.3$ Gyr show a strong trend in line velocity with impact parameter on the scales of the COS-Halos data, because only at these times are the shells small enough that impact parameters of $b=0-150$ kpc probe the majority of the shell structure and projection effects reducing the observed velocity are obvious. At later times, the shells are large and impact parameters of $b=0-150$ kpc probe only a small fraction of the shell structure, so projection effects are less obvious, leading to flat absorption line velocities with impact parameter. This flat trend in absorption line velocity with impact parameter matches the \citet{Keeney2017} data, which show little trend in line velocity over the large impact parameter range probed, $b=0-350$ kpc. The full range of observed absorption line velocities across the many galaxies in the data can be produced by bubbles with different parameters observed at similar bubble ages, or by observing bubbles with similar parameters at many different bubble ages. These two scenarios are impossible to distinguish with absorption studies, particularly because each galaxy typically has only one or two lines of sight passing through its CGM.

\begin{figure*}
\begin{minipage}{175mm}
\centering
\includegraphics[width=\linewidth]{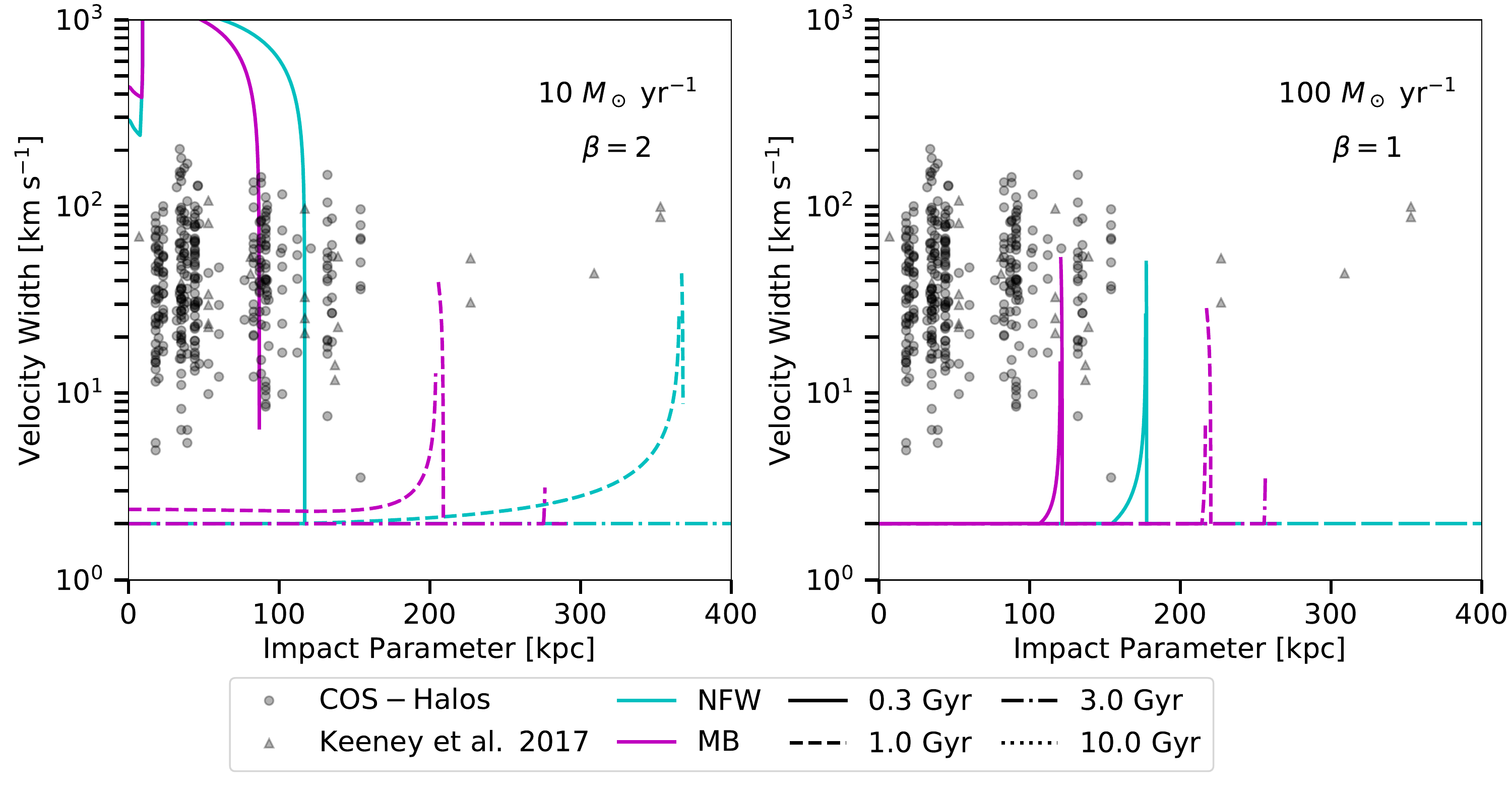}
\caption{The velocity width of the absorption lines for models with $\beta=2$ and SFR $=10\ M_\odot$ yr$^{-1}$ (left panel) or $\beta=1$ and SFR $=100\ M_\odot$ yr$^{-1}$ (right panel) as functions of impact parameter. Curve colors indicate different density profiles for the halo gas, curves with the NFW profile have larger impact parameters than curves with the MB profile}. Curve line style indicates bubble age: $0.3$ Gyr (solid), 1 Gyr (dashed), 3 Gyr (dot-dashed), and 10 Gyr (dotted). Translucent black points indicate observed absorption line widths in the COS-Halos sample (circles) or \citet{Keeney2017} (triangles) for all species of all ionization states.
\label{fig:vel_width}
\end{minipage}
\end{figure*}

Figure~\ref{fig:vel_width} shows the predicted velocity widths of absorption lines for the same subset of models as in Figure~\ref{fig:vcen} (equation~\ref{eq:velwidth}), compared to the velocity FWHM of each component in the \citet{Werk2014} and \citet{Keeney2017} data, which was converted from the reported Doppler broadening values to FWHM by multiplying by a factor of $\sqrt{8\ln 2}$. Because we expect thermal broadening of $\sim2-4$ km s$^{-1}$ for the low-ions at $10^4$ K, we set a lower limit on the velocity widths of absorption lines of 2 km s$^{-1}$. For SFR $=10\ M_\odot$ yr$^{-1}$, the absorption line widths are large, $100-1000$ km s$^{-1}$, when the bubble is 0.3 Gyr old because the wind is still blowing. The absorption line widths drop only for impact parameters that graze the edge of the absorbing region at $b=R_\mathrm{FS}$. These absorption lines are centred at 0 km s$^{-1}$ (see Fig.~\ref{fig:vcen}) and stretch from $-v_\mathrm{wind}$ to $v_\mathrm{wind}$. The absorption velocity widths for bubbles with the power law halo gas density profiles are not plotted, but are similar to the widths in the MB and NFW profiles.

At older bubble ages, when $t>t_\mathrm{SFR}$, the wind has turned off and the only cool gas absorption arises from the cooled shocked wind and cooled shocked swept-up halo gas. The absorption lines are very narrow, no larger than the minimum value we impose from thermal broadening. The predicted absorption line widths increase to match the observed values only for lines of sight with impact parameters that graze the thin shell of the bubble, which are statistically unlikely. Because our model bubble structure predicts such a thin shell, we greatly under-predict the width of absorption lines, despite being consistent with the velocity centres of absorption lines observed.

There are a number of effects we do not model that could increase the width of absorption lines. A variety of halo gas densities along the path of the bubble, such as clumps or clouds, would slow down only the part of the bubble that impacts them, altering the structure of the bubble and breaking spherical symmetry. A line of sight that passes through the more extended, clumpy bubble structure would impact a greater range of gas velocities, producing multiple unresolved absorption components that blend into a broader absorption line. Fragmentation of the bubble's shell due to instabilities (\S\ref{sec:instab}) could also produce a similar effect, and the velocity of instability fingers as they grow at $\sim10-30$ km s$^{-1}$ also broadens the velocity spread of absorption lines early in the shell's evolution while it is unstable. We do not model the evolution of instabilities, but it is possible that the Rayleigh-Taylor fingers form clouds with a spread in velocity that could broaden the absorption lines even after the shell's unstable phase. It is unlikely that the halo gas has the smoothly changing, homogeneous density that we model here, so we do not expect our idealized models to accurately reproduce the absorption line velocity widths observed in nature. Fully hydrodynamic, 3D models are needed to assess the effects of an inhomogenous halo gas and the development of instabilities in the bubble.

The qualitatively similar radial profiles of column density between the different models in \S\ref{sec:den_temp} and the full absorption line velocity parameter space probed by the models shows that it is difficult to predict the properties of the wind or halo gas purely from absorption line observations. Nearly all models are consistent with a majority of the COS-Halos and \citet{Keeney2017} column densities, as the main differences between the models are most evident early in the bubble's evolution, when it has the smallest radius where there are no absorption line data. The drastic evolution through absorption line velocity and impact parameter space as the bubbles age also shows that the data cannot distinguish between the wind parameters driving the bubbles. Distinguishing differences between the models may arise when more physics is included, or in a fully hydrodynamic simulation. However, the success of the model in reproducing the column density and absorption line velocity observations indicates that a supernova-driven wind bubble model presents a consistent picture, even if it is missing more complex physical details.

%%%%%%%%%%%%%%%%%%%%%%%%%%%%%%%%%%%%%%%%%%%%%%%%%%%%%%%%%%%%%%%%%%%%%%%%%%%%%%%%%%%%%%%%%%%%%%%%%%%
\section{Discussion}
\label{sec:discussion}
%%%%%%%%%%%%%%%%%%%%%%%%%%%%%%%%%%%%%%%%%%%%%%%%%%%%%%%%%%%%%%%%%%%%%%%%%%%%%%%%%%%%%%%%%%%%%%%%%%%

%%%%%%%%%%%%%%%%%%%%%%%%%%%%%%%%%%%%%%%%%%%%%%%%%%%%%%%%%%%%%%%%%%%%%%%%%%%%%%%%%%%%%%%%%%%%%%%%%%%
\subsection{Comparison to Simulations}

Cosmological simulations are overwhelmingly in agreement that feedback, whatever its form and driving mechanism, is responsible for many aspects of galaxy evolution and CGM properties. Simulations focus on producing the most realistic galaxies, compared to observations, whereas our analytic model helps develop a physical understanding of the interaction of galactic winds with the CGM and the observational results of different sets of parameters and initial conditions. Many simulations of supernova-driven winds that do not explicitly set the mass loading factor of winds find $\beta\sim0.1-10$, a range that is consistent with the values of $\beta$ that we explore here \citep{Hopkins2012,Sarkar2015}. The idealized simulations of supernova-driven winds by \citet{Fielding2017} most closely match our idealized setup. They find winds reach hundreds of kpc from the host galaxy over several Gyr, in agreement with our results. Without including ionization modeling, we cannot make more detailed comparisons to column densities of various ions that are often calculated. Simulations also find that supernova feedback and radiative cooling are critical to producing cool gas in the galaxy halo that can be traced by low-ionization absorption lines \citep{Hopkins2012,Suarez2016,Biernacki2017,Turner2017}. Despite the idealized nature of our model, it agrees well with many simulations, in addition to observations. We produce similar radial extents of physical densities and radial velocities as \citet{Fielding2017}. Our bubbles travel to hundreds of kpc where they have speeds of $\sim200$ km s$^{-1}$, and produce cool gas in their shells, similar to the findings of \citet{Sarkar2015}. Our model provides a physical framework for understanding the extent and longevity of cool gas in the CGM of simulated galaxies.

Many cosmological simulations \citep[e.g.,][]{Guedes2011,Ford2013,Crain2015} do not directly trace the thermodynamic history of gas in outflows, relying instead on ``subgrid" models to assign properties to the particles that are ejected from the galaxies, so we do not expect wind bubbles to appear in such simulations for comparison to analytic models. However, analytic wind models like the one presented in this paper can be used to inform the evolution of particles in a cosmological simulation, assuming that a particle tracks a fragment of the bubble's shell. In this way, properties such as temperature, mass, and metallicity of particles can be linked to the hydrodynamic model that operates on subgrid scales and a more accurate model that depends on the properties of the galaxy and its wind can be developed.

%%%%%%%%%%%%%%%%%%%%%%%%%%%%%%%%%%%%%%%%%%%%%%%%%%%%%%%%%%%%%%%%%%%%%%%%%%%%%%%%%%%%%%%%%%%%%%%%%%%
\subsection{Other Forms of Feedback}
\label{sec:other_feedback}

We have formulated our model on the idea that supernovae are the only launching mechanism for galactic winds because this assumption gives us a simple relation between properties of the galaxy, like the SFR, and the properties of the wind. However, there are many other forms of star formation feedback that may drive galactic wind bubbles, and our model can be easily adapted to include them by including other momentum and energy source terms. Other forms of star formation feedback include stellar winds, radiation pressure from luminous young stars, cosmic rays, and hot winds driven by Type Ia supernovae in old stellar populations \citep{Ciotti2001}. The term $P_\mathrm{SW}$ in equation~(\ref{eq:energyconserving}) or the term $\dot p_w$ in equation~(\ref{eq:momentumconserving}) can be replaced with pressures and momenta appropriate for different types of feedback, or a sum of several types.

The momentum injection rate on a shell by radiation pressure from a galaxy with luminosity $L_\star$ is $\dot p_\mathrm{rad}\sim L_\star/c$. Comparing to the momentum injection rate for a hot, supernova-driven flow $\dot p_w = (\alpha\beta)^{1/2} (\dot E_w \dot M_w)^{1/2}$ (see \S\ref{sec:eqnsofmotion}), we find
\begin{equation}
\frac{\dot p_w}{L_\star/c} \sim 5 (\alpha\beta)^{1/2} \label{eq:rad_pres}
\end{equation}
where $L_\star\sim10^{11}L_\odot\left(\frac{\mathrm{SFR}}{10\ M_\odot\ \mathrm{yr}^{-1}}\right)$. We can adapt radiation pressure driven outflows into our model by reducing the momentum injection by a factor of 5, which is equivalent to reducing the wind launch velocity to $\sim200$ km s$^{-1}$. In the wind bubble prescription, radiation pressure launches outflows that then shock on the ambient halo gas to produce the hot, shocked wind bubble. We find that radiation pressure driven winds launch galactic wind bubbles that reach a maximum distance of $10-15$ kpc in $100-200$ Myr before falling back to the galaxy, and bubbles driven with larger mass loadings do not travel as far as bubbles driven with smaller mass loadings. The shocked wind radiatively cools in all cases. Because the bubbles do not expand very far into the halo gas, they are not as massive as the bubbles in our fiducial supernova-driven winds model, reaching only $\sim0.03-1.5\times10^{10}M_\odot$. Radiation pressure alone cannot drive galactic wind bubbles to large enough distances to be consistent with the COS-Halos results, but a combination of radiation pressure with other wind driving mechanisms may, as found by \citet{Murray2011,Hopkins2012}.

Active galactic nuclei (AGN) may also launch energy-conserving outflows. \citet{FaucherGiguere2012} developed a wind-blown bubble model for AGN winds, which differ from supernova-driven winds by having a significantly larger launch velocity, $v_l\sim30,000$ km s$^{-1}$. AGN winds are not directly related to the galaxy's SFR and instead the mass outflow rate of the wind is given by $\dot M_\mathrm{AGN}\sim\frac{L_\mathrm{AGN}}{c v_l}$, where we take $L_\mathrm{AGN}\sim10^{46}$ ergs s$^{-1}$ as the AGN luminosity, and $c$ is the speed of light. For $v_l\sim30,000$ km s$^{-1}$, $\dot M_\mathrm{AGN}\sim1.8M_\odot$ yr$^{-1}$. Observations of the quasar proximity effect estimate that AGN are luminous for $\sim1-30$ Myr \citep{Schirber2004}, so we adopt a cutoff time for wind driving of $t_\mathrm{SFR}=20$ Myr. We find that AGN-driven wind bubbles reach very large distances from the galaxy of $R_\mathrm{CD}\sim1000$ kpc over an extended period of time, $\sim10-20$ Gyr. The shocked wind never radiatively cools and the bubble remains hot, so an AGN-driven bubble is prima facie inconsistent with the cool gas traced by low ionization metal absorption lines observed in COS-Halos. However, we do not model Compton cooling, which may be important on small scales in AGN winds \citep{FaucherGiguere2012} and may still allow the hot shocked wind to cool at the very early times we do not consider, so our model may not be predictive in the case of AGN outflows. The AGN bubble remains fast as it reaches hundreds of kpc from the galaxy, at $v_\mathrm{CD}\sim500-700$ km s$^{-1}$, much faster than the velocities of the observed absorption lines. AGN feedback alone may not produce cool gas in the galactic halo, but a combination of AGN and supernova feedback might \citep{Biernacki2017}.

\citet{Farber2017} found that cosmic rays drive gas out of the galactic disk with velocities $v_w\sim200-400$ km s$^{-1}$ and mass loading factors $\beta\sim2-3$. Cosmic rays that are decoupled from the cold ISM gas enhance star formation to $\sim6\ M_\odot$ yr$^{-1}$ for $\sim200$ Myr by allowing cold gas to collapse. We calculated the evolution of a bubble driven by $300$ km s$^{-1}$ wind for 200 Myr, with $\dot M_w=3\times6\ M_\odot$ yr$^{-1}=18\ M_\odot$ yr$^{-1}$, to simulate a galactic wind bubble driven by cosmic rays. The results are qualitatively similar to our fiducial case with SFR $=1\ M_\odot$ yr$^{-1}$ and $\beta=4$ --- the bubble radiatively cools early in its evolution, but the combination of low launch velocity and high $\dot M_w$ keeps it from traveling very far and it reaches only $\sim30$ kpc. Galactic wind bubble driving by cosmic rays alone is therefore not consistent with the COS-Halos observations of cool gas and metals at large impact parameters, but a combination of cosmic ray driving with supernovae, AGN, or radiation pressure may drive a galactic wind bubble further than any wind driving process alone.

No single wind bubble driving mechanism, when modified to fit into our bubble model, produces a large mass of cool gas at large impact parameters that can reproduce the COS-Halos observations like our supernovae-driven fiducial model can.

%%%%%%%%%%%%%%%%%%%%%%%%%%%%%%%%%%%%%%%%%%%%%%%%%%%%%%%%%%%%%%%%%%%%%%%%%%%%%%%%%%%%%%%%%%%%%%%%%%%
\subsection{Infalling Gas}
\label{sec:SFafter}

Throughout this paper, we model the star formation history of the wind-launching galaxy by assuming star formation ceases when $4\times10^9\ M_\odot$ of gas is converted into stars. If additional gas is supplied to the galaxy, it could sustain star formation for an extended period of time. To test how this affects our results, we set SFR $=1\ M_\odot$ yr$^{-1}$ for $t>t_\mathrm{SFR}$. In general, we find this change does not affect which sets of parameters allows the shocked wind to radiatively cool, the thickness of the shocked wind region $R_\mathrm{CD}-R_\mathrm{RS}$ or shocked swept-up halo gas region $R_\mathrm{FS}-R_\mathrm{CD}$, or the column densities of the shocked wind. However, the maximum distance reached by the bubbles for certain sets of parameters increases, and many bubbles that fall back to the galaxy after the wind shuts off instead continue expanding outward at low velocity when star formation is sustained. A late-time hot wind may also be sustained by Type Ia supernovae after star formation ceases \citep{Ciotti2001}, and would produce a similar effect.

As evidenced by Figures~\ref{fig:vcen} and~\ref{fig:vel_width}, the cooled pre-shock wind produces very wide absorption centred at 0 km s$^{-1}$ for any impact parameter $b>r_\mathrm{cool}$ while the wind is still blowing. A continuing low level of star formation after $t_\mathrm{SFR}$ thus produces broad absorption from the cooled shocked wind that is observable even after $t_\mathrm{SFR}$. Sustained star formation produces absorption lines that are too broad, with full widths of $\sim1000$ km s$^{-1}$, similar to the 0.3 Gyr (solid) curves in the left panel of Figure~\ref{fig:vel_width}.

In addition to providing further star formation after $t_\mathrm{SFR}$, infalling gas can directly interact with the outflowing gas. Throughout this paper, we have assumed the halo gas pressure is provided by a static halo. Because the dynamics of the wind bubble are strongly dependent on the halo gas density and pressure, changes to the halo gas, such as infalling or rotating gas, could change our results. We do not model these additional effects.

%%%%%%%%%%%%%%%%%%%%%%%%%%%%%%%%%%%%%%%%%%%%%%%%%%%%%%%%%%%%%%%%%%%%%%%%%%%%%%%%%%%%%%%%%%%%%%%%%%%
\subsection{Multiple Shells from Multiple Starbursts}
\label{sec:multishell}

Although we model a single shell launched at $t=0$ for simplicity, multiple shells could be launched by multiple epochs of star formation. After the first bubble sweeps up the halo gas, the density profile of gas into which each subsequent shell expands would be quite different than the initial density profile of the unperturbed halo gas the first bubble impacted. As the area behind the initial bubble is mostly evacuated, the subsequent bubbles will travel faster than the first, and potentially catch up to it. Assuming no fragmentation of either shell before this occurs, the momentum and energy of each bubble could combine to help push the edge of the initial shell out further. The combined bubble would reach larger distances from the galaxy than a single bubble, and would have larger overall mass.

Recent cosmological simulations \citep{Keller2015,Muratov2015,Feldmann2017} have found that star formation in $L^\star$ progenitor galaxies consists of several strong bursts at high redshift that drive strong outflows \citep[or may recycle back to the galaxy, driving further star formation bursts, e.g.][]{Christensen2016} before settling to a lower, more consistent star formation rate at low redshift. A currently observed low level of star formation does not imply the star formation rate was not higher at some point in the past, so even galaxies that are not currently classified as ``starbursts" could have been starbursts at high redshift. Combined with the long travel time of the bubble's shell into the CGM, the cool CGM observed around low-redshift galaxies may have been produced by an ancient starburst.

We explore galactic wind bubbles driven both by a short, intense star formation period and by a longer, lower-level of star formation. Both types of bubbles could be launched from a galaxy at different points during its evolution, but because radiative cooling of the shocked wind is easier when launched from a high-SFR burst, the cool CGM may have been more likely formed by a high-redshift starburst. The presence of both the low-SFR and high-SFR bubbles in the CGM of the same galaxy could explain some of the multiphase nature of observed CGM gas, but more detailed modeling than we do here is necessary to predict the interaction of both types of bubbles in the CGM.

Another way to produce multiple, interacting bubbles in the CGM is by varying the launch sites not just in time, but also in space. \citet{Schneider2018} examined the interaction of multiple outflowing shells when launched from sites within a galactic disk that shift locations in time. The result was an extended, complex, multiphase outflow structure, but interaction of the outflowing shells with ambient CGM gas was not explored. A full 3D simulation like this is necessary to track the interaction between multiple bubble shells, and further work is needed to follow this outflow structure to larger distances from the galaxy for a detailed comparison to the COS-Halos and \citet{Keeney2017} observational results.

%%%%%%%%%%%%%%%%%%%%%%%%%%%%%%%%%%%%%%%%%%%%%%%%%%%%%%%%%%%%%%%%%%%%%%%%%%%%%%%%%%%%%%%%%%%%%%%%%%%
\subsection{Momentum Boost}

\begin{figure*}
\begin{minipage}{175mm}
\centering
\includegraphics[width=\linewidth]{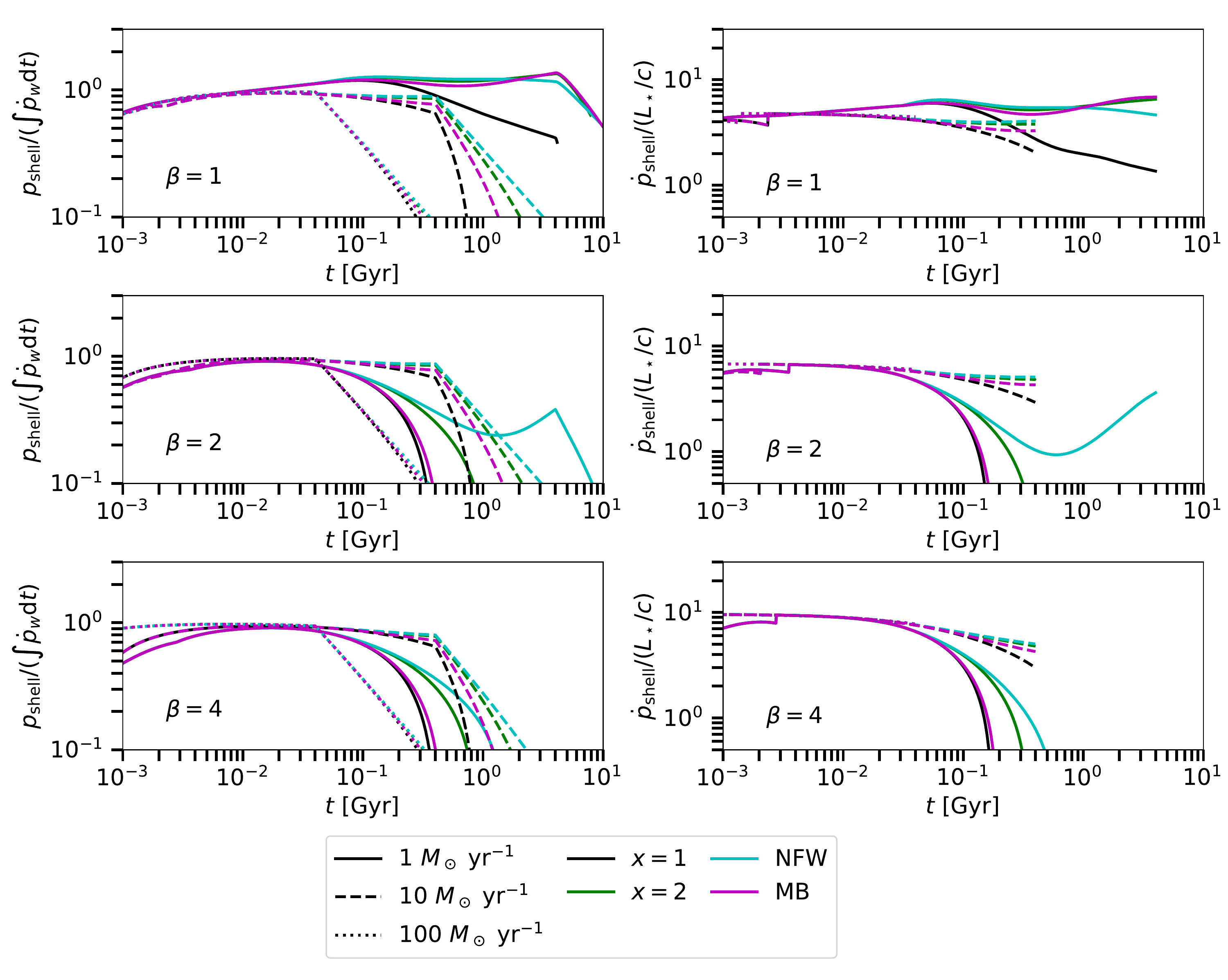}
\caption{Two forms of the momentum boost for all fiducial models as functions of time. Left panels show the momentum boost given by equation~(\ref{eq:boost1}) and right panels show the momentum boost given by equation~(\ref{eq:boost2}). Curve colors and styles indicate halo gas density profiles and star formation rates, as in other figures.}
\label{fig:momboost}
\end{minipage}
\end{figure*}

We calculate the instantaneous ``momentum boost" \citep{FaucherGiguere2012,Richings2017} as the ratio between the bubble's instantaneous momentum and the integrated momentum input of the wind,
\begin{equation}
f_\mathrm{boost}=\frac{p_\mathrm{shell}}{\int \dot p_w\ \mathrm{d}t}. \label{eq:boost1}
\end{equation}
We plot this momentum boost in the left panels in Figure~\ref{fig:momboost}. The models with $\beta=1$ and SFR $=1\ M_\odot$ yr$^{-1}$ never radiatively cool, so these bubbles get a ``boost" of $f_\mathrm{boost}\sim1.5$ from the shock-heating of the wind driving them. The boost comes from the energy-conserving phase of the bubble's evolution, so the bubbles with early radiatively cooling wind do not have instantaneous $f_\mathrm{boost}>1$. The shell's momentum can be smaller than the total time integrated wind momentum when the shell gains enough mass for the gravitational force to be significant, leading to $f_\mathrm{boost}<1$. Including the integrated work done by the bubble produces a larger momentum boost for those bubbles with radiatively cooling shocked wind, but we calculate the instantaneous momentum boost for easier comparison with observations.

Another version of the instantaneous momentum boost, more analogous to active galactic nuclei (AGN) driven winds \citep{Cicone2014}, can be calculated as the ratio of the shell's momentum change to the photon momentum of the starburst\footnote{Note that equation~(\ref{eq:boost2}) is equivalent to equation~(\ref{eq:rad_pres}) when the shocked wind cools early in the evolution of the bubble, because the only momentum input to the shell in that case is the direct momentum injection of the wind. For models without radiatively cooling shocked wind, the shell receives additional momentum input from the energy injection of the wind as well as the wind momentum injection, which is the source of the ``boost."}:
\begin{equation}
f_\mathrm{boost,ph}=\frac{\dot p_\mathrm{shell}}{L_\star/c} \label{eq:boost2}
\end{equation}
where we assume $L_\star\sim10^{11}L_\odot(\mathrm{SFR}/10\ M_\odot\ \mathrm{yr}^{-1})$. We plot this alternate version of the momentum boost in the right panels of Figure~\ref{fig:momboost}. Each bubble's maximum $f_\mathrm{boost,ph}$ is $\sim5-10$, and decreases as the bubble slows down due to gravity. The maximum $f_\mathrm{boost,ph}$ reached depends on $\beta$, but not on SFR or halo gas density profile. Radiatively cooling models with $\beta=1$ have maximum $f_\mathrm{boost,ph}\sim5$, models with $\beta=2$ have maximum $f_\mathrm{boost,ph}\sim7$, and models with $\beta=4$ have maximum $f_\mathrm{boost,ph}\sim10$. The one set of models that does not cool has a slightly larger maximum $f_\mathrm{boost,ph}\sim8$ than the other models with $\beta=1$. Because $\dot p_\mathrm{shell}$ depends more strongly on the mass-loading factor and cooling of the bubble than on the properties of the starburst itself, it is not sufficient to predict the momentum impact of a galactic wind bubble on the CGM by scaling with SFR or $L_\star$ alone.

%%%%%%%%%%%%%%%%%%%%%%%%%%%%%%%%%%%%%%%%%%%%%%%%%%%%%%%%%%%%%%%%%%%%%%%%%%%%%%%%%%%%%%%%%%%%%%%%%%%
\section{Summary}
\label{sec:summary}
%%%%%%%%%%%%%%%%%%%%%%%%%%%%%%%%%%%%%%%%%%%%%%%%%%%%%%%%%%%%%%%%%%%%%%%%%%%%%%%%%%%%%%%%%%%%%%%%%%%

We present a one-dimensional semi-analytic model for a wind bubble blowing out of a galaxy, and compare the predictions of such a model with observations of low-ionization state metal and hydrogen absorption lines toward background quasars from COS-Halos \citep{Werk2014} and \citet{Keeney2017}. We vary the initial conditions, including mass loading, halo gas density profile into which the bubble expands, star formation rate, efficiency of wind heating, and wind launch radius. We focus on the parameter sets that produce cool gas, either in the shocked wind, pre-shock wind, or swept-up halo gas, as low-ionization state metal absorption arises from cool gas. Our main results are as follows:
\begin{enumerate}
\item The halo gas density profile through which the wind bubble expands is the most important factor that determines the distance reached by the shell. For realistic density profiles and default assumptions, shells expand to several hundred to several thousand kiloparsecs from the galaxy over the course of their evolution, regardless of mass-loading $\beta$ or SFR. The large distance reached by the shells is important for predicting the observed absorption at impact parameters $b\sim150-300$ kpc.
\item Bubbles reach large distances slowly, spending $\sim1$ Gyr to $\sim5-10$ Gyr after launch at $\gtrsim100$ kpc from the galaxy. Those bubbles driven by a short starburst reach this distance and remain at least 100 kpc from the galaxy for several Gyr, traveling at low velocities (see (vii) below), even well after star formation has shut off and the host galaxy could be classified as passive. This may explain the otherwise puzzling COS-Halos observations of passive galaxies.
\item All bubble models with either large SFR or large mass-loading $\beta$ produce radiatively cooled shocked wind. The swept-up, shocked halo gas radiatively cools in all our models. We verify that when the free-flowing wind radiatively cools, the shocked wind does as well.
\item The combined mass of wind material and swept-up halo gas in the bubble when it is further than 100 kpc from the galaxy is $\sim2-5\times10^{10}\ M_\odot$ for most fiducial models. In the models with radiatively cooling shocked wind (all but $\beta=1$ and SFR $=1\ M_\odot$ yr$^{-1}$), the total mass of the shell is cool. This is consistent with the strict lower limit estimate of cool gas in the CGM from COS-Halos of $2.1\times10^{10}\ M_\odot$ \citep{Werk2014} and the estimate of cool gas mass from \citet{Keeney2017} of $3.2^{+3.1}_{-1.6}\times10^{10}\ M_\odot$.
\item The effective mass-loading, $\beta_\mathrm{eff}$, of the wind bubbles increases over its original launch value, $\beta$, as the bubbles evolve and sweep up halo gas. By the end of the bubble's evolution, the effective-mass loading is $\sim5-12$, and is similar across all models, despite differences in the mass loading factor at launch (Figure~\ref{fig:betaeff}).
\item Hydrogen and silicon area-averaged column densities from all ionization states are $\log N_\mathrm{H}\sim18-21$ and $\log N_\mathrm{Si}\sim14-16$ over impact parameters $20\lesssim b\lesssim300$ kpc, consistent with the COS-Halos $\log N_\mathrm{H}$ and $\log N_\mathrm{Si}$ measurements and the $\log N_\mathrm{H}$ measurements of \citet{Keeney2017}. There are no large differences in column densities between the models at the impact parameters of the observations.
\item Predicted absorption line velocities cover the full range of $v_\mathrm{cen}\sim-400$ to $400$ km s$^{-1}$ over impact parameters $20\lesssim b \lesssim 300$ kpc observed by COS-Halos and \citet{Keeney2017}, for a variety of wind bubble ages.
\item Absorption line velocity widths we predict are $\sim2-10$ km s$^{-1}$ when the wind is no longer blowing, or $\sim1000$ km s$^{-1}$ when it is still blowing. The wind bubble models either significantly under-predict or significantly over-predict the observed line velocity widths of $\sim5-200$ km s$^{-1}$. The mismatch may be due to the assumption that the halo gas density profile is smooth and spherically symmetric, or the fact that we do not model shell fragmentation due to Vishniac or Rayleigh-Taylor instabilities or multiple interacting shells that could produce unresolved multicomponent absorption from the distribution of velocities that shell fragments would span.
\item We find the momentum boost, calculated as the ratio of the bubble's momentum to the time-integrated wind momentum, to be $\lesssim1$ for all bubbles with radiatively cooling shocked wind. This is due to the fact that the bubble experiences a boost only in the energy-conserving phase of its evolution, and bubbles that radiatively cool do so early in their evolution, and so do not have much time to gain a boost. For the set of models without radiatively cooling shocked wind, the momentum boost is $\sim1.5$.
\end{enumerate}

This analytic, physically-motivated model can reproduce many of the observations of cool gas in the CGM of galaxies across several redshifts, but lacks distinguishing power to determine which set of wind parameters reproduces the observations; instead, nearly all models are consistent with the majority of the observations. The success of the radiatively cooling wind bubble model emphasizes the importance of radiative cooling of the wind, either pre- or post-shock, to produce a large mass of cool, metal-rich gas in the CGM. The fact that the bubble ``hangs" at a large distance from the galaxy for a long time emphasizes that ongoing star formation at the time of observation is not necessary for the presence of wind material in the galactic halo. With an understanding of the physical processes affecting the motion of the bubble and the impact of changes to the initial conditions, a more complete galactic wind bubble model can be built. This model represents a first step in interpreting observational and simulation results, and shows that the physical principles of galactic winds, even in an idealized setting, can explain many aspects of the CGM. Avenues for further study include a multi-dimensional model, more variation of the star formation rate or properties of the galaxy over time, interaction of multiple shells from many epochs of star formation, and a careful analysis of predicted absorption lines with ionization modeling.

%%%%%%%%%%%%%%%%%%%%%%%%%%%%%%%%%%%%%%%%%%%%%%%%%%%%%%%%%%%%%%%%%%%%%%%%%%%%%%%%%%%%%%%%%%%%%%%%%%%
\section*{Acknowledgments}
%%%%%%%%%%%%%%%%%%%%%%%%%%%%%%%%%%%%%%%%%%%%%%%%%%%%%%%%%%%%%%%%%%%%%%%%%%%%%%%%%%%%%%%%%%%%%%%%%%%

The authors thank the anonymous referee for providing helpful suggestions for the improvement of the manuscript. We thank Romeel Dav\'e, Brian A. Keeney, Jessica K. Werk, and Sowgat Muzahid for helpful discussion. CL thanks Adam K. Leroy for interesting ideas for further study. We thank the Simons Foundation for funding the workshop \emph{Galactic Winds; Beyond Phenomenology}, where aspects of this work were conceived. CL \& TAT are supported in part by NSF Grant \#1516967 and NASA ATP 80NSSC18K0526. EQ was supported in part by NSF Grant AST-1715070 and a Simons Investigator Award from the Simons Foundation.

\end{document}